\documentclass[onecolumn]{emulateapj}
\newcommand{\etal}{et al.}
\newcommand{\photo}{{\tt photo}}
\newcommand{\ps}{{\tt PSPhot}}
\newcommand{\kms}{$\rm km\,s^{-1}$}
\begin{document}
\title{The Seventh Data Release of the Sloan Digital Sky Survey}

\author{
Kevork N. Abazajian\altaffilmark{\ref{Maryland}},
Jennifer K. Adelman-McCarthy\altaffilmark{\ref{Fermilab}},
Marcel A. Ag\"ueros\altaffilmark{\ref{Columbia},\ref{NSFFellow}},
Sahar S. Allam\altaffilmark{\ref{Fermilab},\ref{Wyoming}},
Carlos Allende Prieto\altaffilmark{\ref{UCL}},
Deokkeun An\altaffilmark{\ref{OSU},\ref{IPAC}},
Kurt S. J. Anderson\altaffilmark{\ref{APO},\ref{NMSU}},
Scott F. Anderson\altaffilmark{\ref{Washington}},
James Annis\altaffilmark{\ref{Fermilab}},
Neta A. Bahcall\altaffilmark{\ref{Princeton}},
C.A.L. Bailer-Jones\altaffilmark{\ref{MPIA}},
J. C. Barentine\altaffilmark{\ref{Texas}},
Bruce A. Bassett\altaffilmark{\ref{SAAO},\ref{CapeTown}},
Andrew C. Becker\altaffilmark{\ref{Washington}},
Timothy C. Beers\altaffilmark{\ref{MSUJINA}},
Eric F. Bell\altaffilmark{\ref{MPIA}},
Vasily Belokurov\altaffilmark{\ref{Cambridge}}, 
Andreas A. Berlind\altaffilmark{\ref{Vanderbilt}},
Eileen F. Berman\altaffilmark{\ref{Fermilab}},
Mariangela Bernardi\altaffilmark{\ref{Penn}}, 
Steven J. Bickerton\altaffilmark{\ref{Princeton}}, 
Dmitry Bizyaev\altaffilmark{\ref{APO}},
John P. Blakeslee\altaffilmark{\ref{Herzberg}},
Michael R. Blanton\altaffilmark{\ref{NYU}},
John J. Bochanski\altaffilmark{\ref{Washington},\ref{MITKavli}},
William N. Boroski\altaffilmark{\ref{Fermilab}},
Howard J. Brewington\altaffilmark{\ref{APO}},
Jarle Brinchmann\altaffilmark{\ref{Leiden},\ref{Porto}},
J. Brinkmann\altaffilmark{\ref{APO}},
Robert J. Brunner\altaffilmark{\ref{Illinois}},
Tam\'as Budav\'ari\altaffilmark{\ref{JHU}},
Larry N. Carey\altaffilmark{\ref{Washington}},
Samuel Carliles\altaffilmark{\ref{JHU}},
Michael A. Carr\altaffilmark{\ref{Princeton}},
Francisco J. Castander\altaffilmark{\ref{Barcelona}},
David Cinabro\altaffilmark{\ref{WayneState}},
A. J. Connolly\altaffilmark{\ref{Washington}},
Istv\'an Csabai\altaffilmark{\ref{Eotvos}},
Carlos E. Cunha\altaffilmark{\ref{Michigan}},
Paul C. Czarapata\altaffilmark{\ref{Fermilab}},
James R. A. Davenport\altaffilmark{\ref{SanDiego}},
Ernst de Haas\altaffilmark{\ref{PrincetonPhys}},
Ben Dilday\altaffilmark{\ref{ChicagoPhys},\ref{CfCP},\ref{Rutgers}},
Mamoru Doi\altaffilmark{\ref{IoaUT},\ref{IPMU}},
Daniel J. Eisenstein\altaffilmark{\ref{Arizona}},
Michael L. Evans\altaffilmark{\ref{Washington}},
N. W. Evans\altaffilmark{\ref{Cambridge}},
Xiaohui Fan\altaffilmark{\ref{Arizona}},
Scott D. Friedman\altaffilmark{\ref{STScI}},
Joshua A. Frieman\altaffilmark{\ref{Fermilab},\ref{Chicago},\ref{CfCP}},
Masataka Fukugita\altaffilmark{\ref{ICRRUT}},
Boris T. G\"ansicke\altaffilmark{\ref{Warwick}},
Evalyn Gates\altaffilmark{\ref{CfCP}},
Bruce Gillespie\altaffilmark{\ref{JHU}},
G. Gilmore\altaffilmark{\ref{Cambridge}},
Belinda Gonzalez\altaffilmark{\ref{Fermilab}},
Carlos F. Gonzalez\altaffilmark{\ref{Fermilab}},
Eva K. Grebel\altaffilmark{\ref{Heidelberg}},
James E. Gunn\altaffilmark{\ref{Princeton}},
Zsuzsanna Gy\H{o}ry\altaffilmark{\ref{Eotvos}},
Patrick B. Hall\altaffilmark{\ref{York}},
Paul Harding\altaffilmark{\ref{Case}},
Frederick H. Harris\altaffilmark{\ref{NOFS}},
Michael Harvanek\altaffilmark{\ref{Lowell}},
Suzanne L. Hawley\altaffilmark{\ref{Washington}},
Jeffrey J.E. Hayes\altaffilmark{\ref{NASA}},
Timothy M. Heckman\altaffilmark{\ref{JHU}},
John S. Hendry\altaffilmark{\ref{Fermilab}},
Gregory S. Hennessy\altaffilmark{\ref{USNO}},
Robert B. Hindsley\altaffilmark{\ref{NRL}},
J. Hoblitt\altaffilmark{\ref{Hawaii}}, 
Craig J. Hogan\altaffilmark{\ref{Fermilab}},
David W. Hogg\altaffilmark{\ref{NYU}},
Jon A. Holtzman\altaffilmark{\ref{NMSU}},
Joseph B. Hyde\altaffilmark{\ref{Penn}},
Shin-ichi Ichikawa\altaffilmark{\ref{NAOJ}},
Takashi Ichikawa\altaffilmark{\ref{Tohoku}},
Myungshin Im\altaffilmark{\ref{Seoul}}, 
\v{Z}eljko Ivezi\'{c}\altaffilmark{\ref{Washington}},
Sebastian Jester\altaffilmark{\ref{MPIA}},
Linhua Jiang\altaffilmark{\ref{Arizona}},
Jennifer A. Johnson\altaffilmark{\ref{OSU}},
Anders M. Jorgensen\altaffilmark{\ref{NMIMT}},
Mario Juri\'{c}\altaffilmark{\ref{IAS}},
Stephen M. Kent\altaffilmark{\ref{Fermilab}},
R. Kessler\altaffilmark{\ref{CfCP}}, 
S. J. Kleinman\altaffilmark{\ref{Gemini}},
G. R. Knapp\altaffilmark{\ref{Princeton}},
Kohki Konishi\altaffilmark{\ref{ICRRUT},\ref{TokyoPhys}},
Richard G. Kron\altaffilmark{\ref{Fermilab},\ref{Chicago}},
Jurek Krzesinski\altaffilmark{\ref{APO},\ref{MSO}},
Nikolay Kuropatkin\altaffilmark{\ref{Fermilab}},
Hubert Lampeitl\altaffilmark{\ref{Portsmouth}},
Svetlana Lebedeva\altaffilmark{\ref{Fermilab}},
Myung Gyoon Lee\altaffilmark{\ref{Seoul}},
Young Sun Lee\altaffilmark{\ref{MSUJINA}},
R. French Leger\altaffilmark{\ref{Washington}},
S\'ebastien L\'epine\altaffilmark{\ref{AMNH}},
Nolan Li\altaffilmark{\ref{JHU}},
Marcos Lima\altaffilmark{\ref{ChicagoPhys},\ref{CfCP},\ref{Penn}},
Huan Lin\altaffilmark{\ref{Fermilab}},
Daniel C. Long\altaffilmark{\ref{APO}},
Craig P. Loomis\altaffilmark{\ref{Princeton}},
Jon Loveday\altaffilmark{\ref{Sussex}},
Robert H. Lupton\altaffilmark{\ref{Princeton}},
Eugene Magnier\altaffilmark{\ref{Hawaii}},
Olena Malanushenko\altaffilmark{\ref{APO}},
Viktor Malanushenko\altaffilmark{\ref{APO}},
Rachel Mandelbaum\altaffilmark{\ref{IAS},\ref{Hubble}},
Bruce Margon\altaffilmark{\ref{SantaCruz}},
John P. Marriner\altaffilmark{\ref{Fermilab}},
David Mart\'{\i}nez-Delgado\altaffilmark{\ref{IAC}},
Takahiko Matsubara\altaffilmark{\ref{Nagoya}},
Peregrine M. McGehee\altaffilmark{\ref{IPAC}},
Timothy A. McKay\altaffilmark{\ref{Michigan}},
Avery Meiksin\altaffilmark{\ref{Edinburgh}},
Heather L. Morrison\altaffilmark{\ref{Case}},
Fergal Mullally\altaffilmark{\ref{Princeton}},
Jeffrey A. Munn\altaffilmark{\ref{NOFS}},
Tara Murphy\altaffilmark{\ref{Edinburgh},\ref{Sydney}},
Thomas Nash\altaffilmark{\ref{Fermilab}},
Ada Nebot\altaffilmark{\ref{Potsdam}},
Eric H. Neilsen, Jr.\altaffilmark{\ref{Fermilab}},
Heidi Jo Newberg\altaffilmark{\ref{RPI}},
Peter R. Newman\altaffilmark{\ref{APO},\ref{Newman}},
Robert C. Nichol\altaffilmark{\ref{Portsmouth}},
Tom Nicinski\altaffilmark{\ref{Fermilab},\ref{CMCElectronics}},
Maria Nieto-Santisteban\altaffilmark{\ref{JHU}},
Atsuko Nitta\altaffilmark{\ref{Gemini}},
Sadanori Okamura\altaffilmark{\ref{DoAUT}},
Daniel J. Oravetz\altaffilmark{\ref{APO}}, 
Jeremiah P. Ostriker\altaffilmark{\ref{Princeton}},
Russell Owen\altaffilmark{\ref{Washington}},
Nikhil Padmanabhan\altaffilmark{\ref{LBL},\ref{Hubble}},
Kaike Pan\altaffilmark{\ref{APO}},
Changbom Park\altaffilmark{\ref{KIAS}},
George Pauls\altaffilmark{\ref{Princeton}},
John Peoples Jr.\altaffilmark{\ref{Fermilab}},
Will J. Percival\altaffilmark{\ref{Portsmouth}},
Jeffrey R. Pier\altaffilmark{\ref{NOFS}},
Adrian C. Pope\altaffilmark{\ref{Hawaii},\ref{LosAlamos}},
Dimitri Pourbaix\altaffilmark{\ref{Princeton},\ref{Bruxelles}},
Paul A. Price\altaffilmark{\ref{Hawaii}}, 
Norbert Purger\altaffilmark{\ref{Eotvos}},
Thomas Quinn\altaffilmark{\ref{Washington}},
M. Jordan Raddick\altaffilmark{\ref{JHU}},
Paola Re Fiorentin\altaffilmark{\ref{Ljubljana},\ref{MPIA}},
Gordon T. Richards\altaffilmark{\ref{Drexel}},
Michael W. Richmond\altaffilmark{\ref{RIT}},
Adam G. Riess\altaffilmark{\ref{JHU}},
Hans-Walter Rix\altaffilmark{\ref{MPIA}},
Constance M. Rockosi\altaffilmark{\ref{Lick}},
Masao Sako\altaffilmark{\ref{Penn},\ref{Stanford}},
David J. Schlegel\altaffilmark{\ref{LBL}},
Donald P. Schneider\altaffilmark{\ref{PSU}},
Ralf-Dieter Scholz\altaffilmark{\ref{Potsdam}},
Matthias R. Schreiber\altaffilmark{\ref{Valparaiso}},
Axel D. Schwope\altaffilmark{\ref{Potsdam}},
Uro\v{s} Seljak\altaffilmark{\ref{Berkeley},\ref{LBL},\ref{Zurich}},
Branimir Sesar\altaffilmark{\ref{Washington}},
Erin Sheldon\altaffilmark{\ref{Brookhaven},\ref{NYU}},
Kazu Shimasaku\altaffilmark{\ref{DoAUT}},
Valena C. Sibley\altaffilmark{\ref{Fermilab}},
A.E. Simmons\altaffilmark{\ref{APO}},
Thirupathi Sivarani\altaffilmark{\ref{MSUJINA},\ref{Florida}},
J. Allyn Smith\altaffilmark{\ref{APeay}},
Martin C. Smith\altaffilmark{\ref{Cambridge}}, 
Vernesa Smol\v{c}i\'{c}\altaffilmark{\ref{Caltech}},
Stephanie A. Snedden\altaffilmark{\ref{APO}},
Albert Stebbins\altaffilmark{\ref{Fermilab}}
Matthias Steinmetz\altaffilmark{\ref{Potsdam}},
Chris Stoughton\altaffilmark{\ref{Fermilab}},
Michael A. Strauss\altaffilmark{\ref{Princeton}},
Mark SubbaRao\altaffilmark{\ref{Chicago},\ref{Adler}},
Yasushi Suto\altaffilmark{\ref{TokyoPhys}},
Alexander S. Szalay\altaffilmark{\ref{JHU}},
Istv\'an Szapudi\altaffilmark{\ref{Hawaii}},
Paula Szkody\altaffilmark{\ref{Washington}},
Masayuki Tanaka\altaffilmark{\ref{ESO}},
Max Tegmark\altaffilmark{\ref{MIT}},
Luis F.A. Teodoro\altaffilmark{\ref{Glasgow}},
Aniruddha R. Thakar\altaffilmark{\ref{JHU}},
Christy A. Tremonti\altaffilmark{\ref{MPIA}},
Douglas L. Tucker\altaffilmark{\ref{Fermilab}},
Alan Uomoto\altaffilmark{\ref{CarnegieObs}},
Daniel E. Vanden Berk\altaffilmark{\ref{PSU},\ref{StVincent}},
Jan Vandenberg\altaffilmark{\ref{JHU}},
S. Vidrih\altaffilmark{\ref{Heidelberg}},
Michael S. Vogeley\altaffilmark{\ref{Drexel}},
Wolfgang Voges\altaffilmark{\ref{MPIEP}},
Nicole P. Vogt\altaffilmark{\ref{NMSU}},
Yogesh Wadadekar\altaffilmark{\ref{Princeton},\ref{Pune}},
Shannon Watters\altaffilmark{\ref{APO},\ref{PS1}}, 
David H. Weinberg\altaffilmark{\ref{OSU}},
Andrew A. West\altaffilmark{\ref{MITKavli}},
Simon D.M. White\altaffilmark{\ref{MPA}},
Brian C. Wilhite\altaffilmark{\ref{Elmhurst}}, 
Alainna C. Wonders\altaffilmark{\ref{JHU}},
Brian Yanny\altaffilmark{\ref{Fermilab}},
D. R. Yocum\altaffilmark{\ref{Fermilab}},
Donald G. York\altaffilmark{\ref{Chicago},\ref{EFI}},
Idit Zehavi\altaffilmark{\ref{Case}},
Stefano Zibetti\altaffilmark{\ref{MPIA}}, 
Daniel B. Zucker\altaffilmark{\ref{Cambridge}}
}

\altaffiltext{1}{
Department of Physics, University of Maryland,  
College Park, MD 20742
\label{Maryland}}

\altaffiltext{2}{
Fermi National Accelerator Laboratory, P.O. Box 500, Batavia, IL 60510.
\label{Fermilab}}

\altaffiltext{3}{
 	Columbia Astrophysics Laboratory, 550 West 120th Street, New
	York, NY 10027.\label{Columbia}}

\altaffiltext{4}{
         NSF Astronomy and Astrophysics
	 Postdoctoral Fellow.\label{NSFFellow}}

\altaffiltext{5}{
Department of Physics and Astronomy, University of Wyoming, Laramie, WY 82071.
\label{Wyoming}}

\altaffiltext{6}{
        Mullard Space Science Laboratory, University College London,
        Holmbury St. Mary, Surrey RH5 6NT, United Kingdom
\label{UCL}}

\altaffiltext{7}{
Department of Astronomy, 
Ohio State University, 140 West 18th Avenue, Columbus, OH 43210.
\label{OSU}}

\altaffiltext{8}{
IPAC, MS 220-6, California Institute of Technology,
Pasadena, CA 91125.\label{IPAC}}

\altaffiltext{9}{
Apache Point Observatory, P.O. Box 59, Sunspot, NM 88349.
\label{APO}}

\altaffiltext{10}{
Department of Astronomy, MSC 4500, New Mexico State University,
P.O. Box 30001, Las Cruces, NM 88003.
\label{NMSU}}

\altaffiltext{11}{
Department of Astronomy, University of Washington, Box 351580, Seattle, WA
98195.
\label{Washington}}

\altaffiltext{12}{
Department of Astrophysical Sciences, Princeton University, Princeton, NJ
08544.
\label{Princeton}}

\altaffiltext{13}{
Max-Planck-Institut f\"ur Astronomie, K\"onigstuhl 17, D-69117 Heidelberg,
Germany.
\label{MPIA}}

\altaffiltext{14}{
McDonald Observatory and Department of Astronomy, The University of
Texas, 1 University Station, C1400,  Austin, TX
78712-0259.\label{Texas}}

\altaffiltext{15}{
South African Astronomical Observatory, Observatory,
  Cape Town, South Africa.\label{SAAO}}

\altaffiltext{16}{
University of Cape Town, Rondebosch, Cape Town, South
  Africa.
\label{CapeTown}}

\altaffiltext{17}{
Dept. of Physics \& Astrophysics, CSCE: Center for the Study of Cosmic 
Evolution, and JINA: Joint Institute for Nuclear Astrophysics, Michigan 
State University, E. Lansing, MI  48824.
\label{MSUJINA}}

\altaffiltext{18}{
Institute of Astronomy, University of Cambridge, Madingley Road,
Cambridge CB3 0HA, UK.
\label{Cambridge}}

\altaffiltext{19}{
Department of Physics and Astronomy, Vanderbilt University, Nashville  
TN 37235.\label{Vanderbilt}}

\altaffiltext{20}{
Department of Physics and Astronomy, University of Pennsylvania,
209 South 33rd Street, Philadelphia, PA 19104. 
\label{Penn}}

\altaffiltext{21}{
Herzberg Institute of Astrophysics,
National Research Council of Canada,
5071 West Saanich Road,
Victoria, B.C.  V9E 2E7.
\label{Herzberg}}

\altaffiltext{22}{
Center for Cosmology and Particle Physics,
Department of Physics,
New York University,
4 Washington Place,
New York, NY 10003.
\label{NYU}}

\altaffiltext{23}{
MIT Kavli Institute for Astrophysics and Space Research,
77 Massachusetts Avenue,
Cambridge, MA 02139
\label{MITKavli}}

\altaffiltext{24}{
  Leiden Observatory, Leiden
  University, PO Box 9513, 2300 RA Leiden, the Netherlands
\label{Leiden}}

\altaffiltext{25}{
Centro de Astrof{\'\i}sica da Universidade do Porto, Rua 
das Estrelas - 4150-762 Porto, Portugal.
\label{Porto}}

\altaffiltext{26}{
Department of Astronomy,
University of Illinois,
1002 West Green Street, Urbana, IL 61801.
\label{Illinois}}

\altaffiltext{27}{
Center for Astrophysical Sciences, Department of Physics and Astronomy, Johns
Hopkins University, 3400 North Charles Street, Baltimore, MD 21218. 
\label{JHU}}

\altaffiltext{28}{
Institut de Ci\`encies de l'Espai (IEEC/CSIC),
  Campus UAB, E-08193  
Bellaterra, Barcelona, Spain.
\label{Barcelona}}

\altaffiltext{29}{
Department of Physics and Astronomy, Wayne 
State University, Detroit, MI USA 48202.
\label{WayneState}}

\altaffiltext{30}{
Department of Physics of Complex Systems, 
E\"{o}tv\"{o}s Lor\'and University, Pf.\ 32,
H-1518 Budapest, Hungary.
\label{Eotvos}}

\altaffiltext{31}{
Departments of Physics and Astronomy, University of Michigan, 450
Church Street, Ann
Arbor, MI 48109.
\label{Michigan}}

\altaffiltext{32}{
Department of Astronomy, San Diego State University, PA 221, 5500
Campanile Drive, San Diego, CA 92182-1221
\label{SanDiego}}

\altaffiltext{33}{
Joseph Henry Laboratories, Princeton University, Princeton, NJ
08544.
\label{PrincetonPhys}}

\altaffiltext{34}{
Department of Physics, University of Chicago, 5640 South
Ellis Avenue, Chicago, IL 60637.
\label{ChicagoPhys}}

\altaffiltext{35}{
Kavli Institute for Cosmological Physics, The University of Chicago,
5640 South Ellis Avenue, Chicago, IL 60637.
\label{CfCP}}

\altaffiltext{36}{Department of Physics and Astronomy
Rutgers, the State University of New Jersey
136 Frelinghuysen Road
Piscataway, NJ 08854-8019
\label{Rutgers}}

\altaffiltext{37}{
Institute of Astronomy,
Graduate School of Science, The University of Tokyo,
2-21-1 Osawa, Mitaka, 181-0015, Japan.
\label{IoaUT}}

\altaffiltext{38}{
Institute for the Physics and Mathematics of the Universe,
The University of Tokyo,
5-1-5 Kashiwanoha, Kashiwa, 277-8568, Japan
\label{IPMU}}

\altaffiltext{39}{
Steward Observatory, 933 North Cherry Avenue, Tucson, AZ 85721.
\label{Arizona}}

\altaffiltext{40}{
Space Telescope Science Institute, 3700 San Martin Drive, Baltimore, MD
21218.
\label{STScI}}

\altaffiltext{41}{
Department of Astronomy and Astrophysics, University of Chicago, 5640 South
Ellis Avenue, Chicago, IL 60637.
\label{Chicago}}

\altaffiltext{42}{
Institute for Cosmic Ray Research, The University of Tokyo, 5-1-5 Kashiwa,
 Kashiwa City, Chiba 277-8582, Japan.
\label{ICRRUT}}

\altaffiltext{43}{
Department of Physics, University of Warwick,
Coventry CV4 7AL, United Kingdom.\label{Warwick}}

\altaffiltext{44}{
Astronomisches Rechen-Institut, Zentrum f\"ur Astronomie,
University of Heidelberg, M\"onchhofstrasse 12-14,
D-69120 Heidelberg, Germany.\label{Heidelberg}}

\altaffiltext{45}{
Dept. of Physics \& Astronomy,
York University,
4700 Keele St.,
Toronto, ON, M3J 1P3,
Canada
\label{York}}

\altaffiltext{46}{
Department of Astronomy, Case Western Reserve University,
Cleveland, OH 44106.
\label{Case}}

\altaffiltext{47}{
US Naval Observatory, 
Flagstaff Station, 10391 W. Naval Observatory Road, Flagstaff, AZ
86001-8521.
\label{NOFS}}

\altaffiltext{48}{
Lowell Observatory, 
1400 W Mars Hill Rd, 
Flagstaff AZ 86001.\label{Lowell}}

\altaffiltext{49}{
Heliophysics Division, Science Mission Directorate, 
NASA Headquarters,
300 E Street SW, Washington DC 20546-0001
\label{NASA}}

\altaffiltext{50}{
US Naval Observatory, 3540 Massachusetts Avenue NW, Washington, DC 20392.
\label{USNO}}

\altaffiltext{51}{
Code 7215, Remote Sensing Division,
Naval Research Laboratory, 
4555 Overlook Avenue SW,
Washington, DC 20392.
\label{NRL}}

\altaffiltext{52}{
Institute for Astronomy, 2680 Woodlawn Road, Honolulu, HI 96822.
\label{Hawaii}}

\altaffiltext{53}{
National Astronomical Observatory, 
2-21-1 Osawa, Mitaka, Tokyo 181-8588, Japan.
\label{NAOJ}}

\altaffiltext{54}{
Astronomical Institute, Tohoku University,
Aoba, Sendai 980-8578, Japan
\label{Tohoku}}

\altaffiltext{55}{
Department of Physics \& Astronomy,
Seoul National University,
Shillim-dong, San 56-1, Kwanak-gu,
Seoul 151-742, Korea
\label{Seoul}}

\altaffiltext{56}{
Electrical Engineering Department,
New Mexico Institute of Mining and Technology,
801 Leroy Place,
Socorro, NM 87801.\label{NMIMT}}

\altaffiltext{57}{
Institute for Advanced Study,
Einstein Drive,
Princeton, NJ 08540.
\label{IAS}}

\altaffiltext{58}{
Gemini Observatory, 670 N. A'ohoku Place, Hilo, HI 96720.
\label{Gemini}}

\altaffiltext{59}{
Department of Physics and Research Center for the Early Universe,
Graduate School of Science, The University of Tokyo, 7-3-1 Hongo, Bunkyo, 
Tokyo 113-0033, Japan.
\label{TokyoPhys}}

\altaffiltext{60}{
Obserwatorium Astronomiczne na Suhorze, Akademia Pedogogiczna w
Krakowie, ulica Podchor\c{a}\.{z}ych 2,
PL-30-084 Krac\'ow, Poland.
\label{MSO}}

\altaffiltext{61}{
Institute of Cosmology and Gravitation (ICG),
Mercantile House, Hampshire Terrace,
Univ. of Portsmouth, Portsmouth, PO1 2EG, UK.
\label{Portsmouth}}

\altaffiltext{62}{
  Department of Astrophysics,
American Museum of Natural History,
Central Park West at 79th Street,
New York, NY 10024\label{AMNH}}

\altaffiltext{63}{
Astronomy Centre, University of Sussex, Falmer, Brighton BN1 9QH, UK. 
\label{Sussex}}

\altaffiltext{64}{
Hubble Fellow.\label{Hubble}}

\altaffiltext{65}{
Department of Astronomy \& Astrophysics, University of California, Santa
Cruz, CA 95064.
\label{SantaCruz}}

\altaffiltext{66}{
Instituto de Astrof\'\i{}sica de Canarias, E38205 La Laguna, Tenerife, Spain.
\label{IAC}}

\altaffiltext{67}{
Department of Physics and Astrophysics,
 Nagoya University,
 Chikusa, Nagoya 464-8602,
 Japan.
\label{Nagoya}}

\altaffiltext{68}{
SUPA, Institute for Astronomy,
Royal Observatory,
University of Edinburgh,
Blackford Hill,
Edinburgh EH9 3HJ,
UK.
\label{Edinburgh}}

\altaffiltext{69}{
Sydney Institute of Astronomy, The University of Sydney,
NSW 2006, Australia
\label{Sydney}}

\altaffiltext{70}{
Astrophysical Institute Potsdam, An der Sternwarte 16, 
14482 Potsdam, Germany.
\label{Potsdam}}

\altaffiltext{71}{
Department of Physics, Applied Physics, and Astronomy, Rensselaer
Polytechnic Institute, 110 Eighth Street, Troy, NY 12180. 
\label{RPI}}

\altaffiltext{72}{
322 Fulham Palace Road, London, SW6 6HS United Kingdom
\label{Newman}}

\altaffiltext{73}{
    CMC Electronics Aurora,
 84 N. Dugan Rd.
    Sugar Grove, IL 60554.
\label{CMCElectronics}}

\altaffiltext{74}{
Department of 
Astronomy and Research Center for the Early Universe, 
Graduate School of Science, The University of Tokyo,
 7-3-1 Hongo, Bunkyo, Tokyo 113-0033, Japan.
\label{DoAUT}}

\altaffiltext{75}{
Lawrence Berkeley National Laboratory, One Cyclotron Road,
Berkeley, CA 94720.
\label{LBL}}

\altaffiltext{76}{
Korea Institute for Advanced Study,
87 Hoegiro, Dongdaemun-Gu, 
Seoul 130-722, Korea
\label{KIAS}}

\altaffiltext{77}{
Los Alamos National Laboratory, PO Box 1663, Los Alamos, NM 87545
\label{LosAlamos}}

\altaffiltext{78}{
FNRS
Institut  d'Astronomie et d'Astrophysique,
 Universit\'e Libre de Bruxelles, CP. 226, Boulevard du Triomphe, B-1050
 Bruxelles, Belgium.
\label{Bruxelles}}

\altaffiltext{79}{
Department of Physics, University of Ljubljana, Jadranska 19, 1000 Ljubljana, Slovenia
\label{Ljubljana}}

\altaffiltext{80}{
Department of Physics, 
Drexel University, 3141 Chestnut Street, Philadelphia, PA 19104.
\label{Drexel}}

\altaffiltext{81}{
Department of Physics, Rochester Institute of Technology, 84 Lomb Memorial
Drive, Rochester, NY 14623-5603.
\label{RIT}}

\altaffiltext{82}{
UCO/Lick Observatory, University of California, Santa Cruz, CA 95064.
\label{Lick}}

\altaffiltext{83}{
   Kavli Institute for Particle Astrophysics \& Cosmology,
   Stanford University, P.O. Box 20450, MS29,
   Stanford, CA 94309.\label{Stanford}}

\altaffiltext{84}{
Department of Astronomy and Astrophysics, 525 Davey Laboratory, 
Pennsylvania State
University, University Park, PA 16802.
\label{PSU}}

\altaffiltext{85}{
Universidad de Valparaiso, Departamento de Fisica y
  Astronomia,
 Valparaiso, Chile.\label{Valparaiso}}

\altaffiltext{86}{
Physics Department, University of
California, Berkeley, CA 94720.
\label{Berkeley}}

\altaffiltext{87}{
Institute for Theoretical Physics, University of Zurich, Zurich 8057  
Switzerland.\label{Zurich}}

\altaffiltext{88}{
Bldg 510
Brookhaven National Laboratory
Upton NY,  11973
\label{Brookhaven}}

\altaffiltext{89}{
Department of Astronomy, University of Florida,
                Bryant Space Science Center, Gainesville, FL
		32611-2055
\label{Florida}}

\altaffiltext{90}{
Department of Physics and Astronomy, Austin Peay State University,
P.O. Box 4608, Clarksville, TN 37040.
\label{APeay}}

\altaffiltext{91}{
California Institute of Technology, 1200 East California Blvd, Pasadena, CA
91125
\label{Caltech}}

\altaffiltext{92}{
Adler Planetarium and Astronomy Museum,
1300 Lake Shore Drive,
Chicago, IL 60605.
\label{Adler}}

\altaffiltext{93}{
European Southern Observatory,
Karl-Schwarzschild-Str. 2,
D-85748 Garching bei M\"{u}nchen, Germany
\label{ESO}}

\altaffiltext{94}{
Dept. of Physics, Massachusetts Institute of Technology, Cambridge,  
MA 02139.
\label{MIT}}

\altaffiltext{95}{
Astronomy and Astrophysics Group,
Department of Physics and Astronomy,
Kelvin Building,
University of Glasgow,
Glasgow G12 8QQ,
Scotland, UK
\label{Glasgow}}

\altaffiltext{96}{
Observatories of the Carnegie Institution of Washington, 
813 Santa Barbara Street, 
Pasadena, CA  91101.
\label{CarnegieObs}}

\altaffiltext{97}{
Department of Physics, Saint Vincent College, 300 Fraser Purchase
Road, Latrobe, PA 15650
\label{StVincent}}

\altaffiltext{98}{
Max-Planck-Institut f\"ur extraterrestrische Physik, 
Giessenbachstrasse 1, D-85741 Garching, Germany.
\label{MPIEP}}

\altaffiltext{99}{
National Centre for Radio Astrophysics, Tata 
Institute of Fundamental Research, Post Bag 3, Ganeshkhind, Pune 411007, 
India
\label{Pune}}

\altaffiltext{100}{
Advanced Technology and Research Center, 
Institute for Astronomy, 34 Ohia Ku St., Pukalani, Hawaii 96768.
\label{PS1}}

\altaffiltext{101}{
Max-Planck-Institut f\"ur Astrophysik, Postfach 1, 
D-85748 Garching, Germany.
\label{MPA}}

\altaffiltext{102}{
Department of Physics,
Elmhurst College,
190 Prospect Ave.,
Elmhurst, IL 60126
\label{Elmhurst}}

\altaffiltext{103}{
Enrico Fermi Institute, University of Chicago, 5640 South Ellis Avenue,
Chicago, IL 60637.
\label{EFI}}

\shorttitle{SDSS DR7}
\shortauthors{Abazajian \etal}

\begin{abstract}
This paper describes the Seventh Data Release of the Sloan Digital Sky
Survey (SDSS), marking the completion of the original goals of the
SDSS and the end of the phase known as SDSS-II.  It
includes 11663 deg$^2$ of imaging data, with most of the $\sim 2000$
deg$^2$ increment over the previous data release lying in regions of
low Galactic latitude. The catalog contains five-band photometry for 357 million
distinct objects.  The survey also includes repeat photometry on a
120$^\circ$ long, 2.5$^\circ$ wide stripe along the Celestial Equator
in the Southern Galactic Cap, with some regions covered by as many as
90 individual imaging runs.  We include a coaddition of the best of these data, going
roughly two magnitudes fainter than the main survey over 250 deg$^2$.  The
survey has completed spectroscopy over 9380 deg$^2$; the spectroscopy
is now complete over a large contiguous area of the Northern Galactic
Cap, closing the gap that was present in previous data releases.
There are over 1.6 million spectra in total, including 930,000
galaxies, 120,000 quasars, and 460,000 stars.

   The data release includes improved stellar photometry at low
   Galactic latitude.  
The astrometry has all been recalibrated with 
  the second version of the USNO CCD Astrograph Catalog (UCAC-2),
  reducing the rms statistical errors at the 
  bright end to 45 milli-arcseconds per coordinate.  We further quantify a
  systematic error in bright galaxy photometry due to poor sky
  determination; this problem is less severe than previously 
  reported for the majority of galaxies.  Finally, we describe a
  series of improvements to the spectroscopic reductions, including
  better flat-fielding and improved wavelength calibration at the blue
  end, better processing of objects with extremely strong narrow emission
  lines, and an improved determination of stellar metallicities. 

\end{abstract}
\keywords{Atlases---Catalogs---Surveys}
\section{Overview of the Sloan Digital Sky Survey}
\label{sec:introduction}
The Sloan Digital Sky Survey (SDSS; York \etal\ 2000) saw first light
a decade ago, with the goals of obtaining CCD imaging in five broad
bands over 10,000 deg$^2$ of high-latitude sky, and spectroscopy of a
million galaxies and one hundred thousand quasars over this same
region.  With this, its seventh public data release, these goals have
been realized.  The survey facilities have also been used to carry out
a comprehensive imaging and spectroscopic survey to explore the
structure, composition, and kinematics of the Milky Way Galaxy (Sloan
Extension for Galactic Understanding and Exploration; SEGUE; Yanny
\etal\ 2009), and a
repeat imaging survey that has discovered almost 500 spectroscopically
confirmed Type Ia supernovae with superb light curves (Frieman \etal\
2008; Holtzman \etal\ 2008).

  The SDSS uses a dedicated wide-field 2.5m telescope (Gunn \etal\
  2006) located at Apache Point Observatory (APO) near Sacramento Peak in
  Southern New Mexico.  The telescope uses two instruments.  The first
  is a wide-field imager (Gunn \etal\ 1998) with 24 $2048 \times 2048$
  CCDs on the focal plane with $0.396^{\prime\prime}$ pixels, that
  covers the sky in drift scan mode in five filters in the order
  $riuzg$ (Fukugita \etal\ 1996).  The imaging is done with the
  telescope tracking great circles at the sidereal rate; the effective
  exposure time per filter is 54.1 seconds, and 18.75 deg$^2$ are 
  imaged per hour in each of the five filters.  The images are mostly
  taken under good seeing conditions (the median is about $1.4''$ in
  $r$) on moonless photometric nights (Hogg \etal\ 2001); the
  exceptions are a series of repeat scans of the Celestial Equator in
  the Fall for a supernova search (Frieman \etal\ 2008), as is
  described in more detail in \S~\ref{sec:stripe82}.  The 95\%
  completeness limits of the images are $u,g,r,i,z = 22.0, 22.2,
  22.2, 21.3, 20.5$, respectively (Abazajian \etal\ 2004), although these values 
 depend as expected on seeing and sky brightness.  The images
  are processed through a series of pipelines that determine an
  astrometric calibration (Pier \etal\ 2003) and detect and measure
  the brightnesses, positions and shapes of objects (Lupton \etal\
  2001; Stoughton \etal\ 2002).  The astrometry is good to 45
  milli-arcseconds (mas) rms
  per coordinate at the bright end, as described in more detail in
  \S~\ref{sec:astrometry}.  The photometry is calibrated to an AB
  system (Oke \& Gunn 1983), and the zeropoints of the system are
  known to 1--2\% (Abazajian \etal\ 2004).  The photometric calibration
  is done in two ways, by tying to photometric standard stars (Smith
  \etal\ 2002) measured
  by a separate 0.5-m telescope on the site (Tucker \etal\ 2006;
  Ivezi\'c \etal\ 2004), and by using the overlap between adjacent imaging
  runs to tie the photometry of all the imaging observations together, in a
  process called 
  ubercalibration (Padmanabhan \etal\ 2008).  Results of both
  processes are made available; with this data release, the
  ubercalibration results, which are uncertain at the $\sim 1\%$
  level in $griz$ and 2\% in $u$, are now the default photometry made
  available in the data release described in this paper.

  The photometric catalogs of detected objects are used to identify
  objects for spectroscopy with the second of the
  instruments on the telescope: a 640-fiber-fed pair of multi-object
  double spectrographs, giving coverage from 3800\AA\ to 9200\AA\ at a 
  resolution of $\lambda/\Delta \lambda \simeq 2000$.  The objects chosen for
  spectroscopic follow-up are selected based on photometry corrected
  for Galactic extinction following Schlegel, Finkbeiner, \& Davis
  (1998; hereafter SFD), and include:
\begin{itemize} 
\item A sample of galaxies complete to a Petrosian (1976) magnitude
  limit of $r=17.77$ (Strauss \etal\ 2002);
\item Two deeper samples of luminous red ellipticals selected in
  color-magnitude space to $r = 19.2$ and $r=19.5$, respectively,
  which produce an approximately volume-limited sample to $z = 0.38$, and
  a flux-limited sample extending to $z=0.55$, respectively
  (Eisenstein \etal\ 2001);
\item Flux limited samples of quasar candidates, selected by their
  non-stellar colors or FIRST (Becker \etal\ 1995) radio emission to
  $i=19.1$ in regions of color space characteristic of $z < 3$
  quasars, and to $i=20.2$ for quasars with $3 < z < 5.5$ (Richards
  \etal\ 2002);
\item A variety of ancillary samples, including optical
  counterparts to ROSAT-detected X-ray sources (Anderson \etal\ 2007);
\item Stars for spectrophotometric calibration and telluric
  absorption correction, as well as regions of blank sky for accurate
  sky subtraction;
\item A variety of categories of stellar targets with a series of color
  and magnitude cuts for measurements of radial velocity,
  metallicity, surface temperature, and Galactic structure as part of 
  SEGUE (Yanny \etal\ 2009).  
\end{itemize}
These targets are arranged on {\em tiles} of radius $1.49^\circ$, with 
centers chosen to maximize the number of targeted objects
(Blanton \etal\ 2003).  Each tile contains 640 objects, and forms the
template for an aluminum spectroscopic plate, in which 
holes are drilled to hold optical fibers that feed the spectrographs.
Spectroscopic exposures are 15 minutes long, and three or
more are taken for a given plate to reach pre-defined requirements of
signal-to-noise ratio (S/N), namely $(S/N)^2 > 15$ per 1.5\AA\ pixel
for stellar 
objects of fiber magnitude $g = 20.2, r=20.25$ and
$i=19.9$.  For the SEGUE faint plates, the
exposures are considerably deeper, and typically consist of eight
15-minute exposures, giving $(S/N)^2 \sim 100$ at the same depth
(Yanny \etal\ 2009).   

  Spectra are extracted and calibrated in wavelength and flux.  The
  typical S/N of a galaxy near the main sample flux limit is 10 per
  pixel.  The broad-band spectrophotometric calibration is accurate to 
  4\% rms for point sources (Adelman-McCarthy \etal\ 2008), and the
  wavelength calibration is 
  good to 2 \kms.  The spectra are classified and
  redshifts determined using a pair of pipelines (Stoughton \etal\ 2002; Subbarao
  \etal\ 2002), which give
  consistent results 98\% of the time; the discrepant objects tend to
  be of very low S/N, or very unusual objects, such as extreme BALs,
  superposed sources, and so on. The vast majority of
  the spectra of galaxies and quasars yield reliable redshifts; the
  failure rate is of order 1\% for galaxies, and slightly larger for
  quasars.  The stellar targets are 
  further processed by a separate pipeline (Lee \etal\ 2008a,b;
  Allende Prieto \etal\ 2008a) which
  determines surface temperatures, metallicities, and gravities.  

The resulting catalogs are stored and distributed via a database
accessible on the web (the Catalog Archive Server, CAS\footnote{\tt
  http://cas.sdss.org/astro}; Thakar \etal\ 2008), and the images and flat
files are available in bulk through the Data Archive
Server (DAS)\footnote{\tt http://das.sdss.org}.  

  The SDSS saw first light in May 1998, and started routine operations
  in April 2000.  It was originally funded for five years of
  operations, but had not completed its core goals of imaging and
  spectroscopy of a large contiguous area of the Northern Galactic Cap
  by 2005.  The survey was extended for 
  an additional three years, with the additional goals of the SEGUE
  and the supernova surveys mentioned above.  The extended program is
  known as SDSS-II, and the component of SDSS-II that represents
  the completion of SDSS-I is known as the Legacy Survey.  SDSS-II
  observations were completed in July 2008.

  The SDSS data have been made public in a series of yearly data
  releases (Stoughton \etal\ 2002; Abazajian \etal\ 2003, 2004, 2005, 2006;
  Adelman-McCarthy \etal\ 2007, 2008; hereafter the EDR, DR1, DR2,
  DR3, DR4, DR5, and DR6 papers, respectively).  The most recent of these
  papers described
  the Sixth Data Release (DR6), which included data taken through July
  2006.  The present paper describes the Seventh Data Release (DR7),
  including data taken through the end of SDSS-II in 2008 July, and
  thus represents two additional years of data.  The data releases are
  cumulative; DR7 includes all data included in the previous releases
  as well.  In
  \S~\ref{sec:sky_coverage}, we describe the footprint of this survey;
  most importantly, we have completed our goals of:
\begin{itemize} 
\item contiguous imaging and spectroscopy over 7500 deg$^2$ of the
  Northern Galactic Cap (the Legacy survey); 
\item imaging and spectroscopy of stellar sources over an additional
  3500 deg$^2$ at  lower Galactic latitudes to study the structure of the Milky
  Way, and
\item repeat imaging of $>250$ deg$^2$ on the Celestial Equator in the
  Fall months to discover Type Ia supernovae with $0.1 < z < 0.4$.  
\end{itemize}

In \S~\ref{sec:runsDB}, we describe the repeat scans on the Celestial
Equator, including a co-addition of the images to reach about two
magnitudes deeper than the main survey.  In \S~\ref{sec:imaging}, we
present improvements in the processing of the imaging data, including
improved stellar photometry at low Galactic latitudes, an astrometric
recalibration, and improvements in our photometric redshift
algorithms for galaxies.  The DR6 paper described a problem with the
photometry of bright galaxies; we explore this further in
\S~\ref{sec:skysub}. In \S~\ref{sec:spectroscopy}, we discuss
improvements in the spectroscopic processing of the data.  The DR6
paper described 
  improvements in the wavelength and spectrophotometric calibration;
  we have implemented further refinements which are important in the
  determination of accurate stellar parameters from the spectra. 

  We conclude in \S~\ref{sec:conclusions} with a discussion of the
  future of the SDSS project. 

\section{Survey Footprint}
\label{sec:sky_coverage}
Table~\ref{table:dr7_contents} summarizes the contents of DR7, giving
the imaging and spectroscopic sky coverage and number of objects. The imaging footprint has increased by 
roughly 22\% since DR6 (most of it outside the contiguous area of the
North Galactic Cap), and the number of spectra has increased by 29\%.

\begin{deluxetable}{lr}
\tablecaption{Coverage and Contents of DR7
              \label{table:dr7_contents}}

\startdata

\cutinhead{\bf Imaging} 

 Imaging area in CAS &11663\ deg$^2$\\
 Imaging catalog in CAS & 357 million unique objects\\
 \quad Legacy footprint area & 8423\ deg$^2$\\
                             & (7646 deg$^2$ in N. Galactic Cap)\\
 \quad Legacy imaging catalog & 230 million unique objects \\
                              & 585 million entries (including duplicates)\\
 \quad SEGUE footprint area, available in DAS\tablenotemark{a} & 3500\
 deg$^2$ (more than double DR6)\\
 \quad SEGUE footprint area, available in CAS & 3240\ deg$^2$\\
 SEGUE imaging catalog & 127 million unique objects\\
 M31, Perseus, Sagittarius scan area & $\sim 46$ deg$^2$\\
 Southern Equatorial Stripe with $> 70$ repeat scans& $\sim 250$ deg$^2$\\
 Commissioning (``Orion'') data &832 deg$^2$\\
\cutinhead{\bf Spectroscopy}
 Spectroscopic footprint area &9380 deg$^2$\\
 \quad Legacy &8032 deg$^2$\\
 \quad SEGUE  &1348 deg$^2$\\
 Total number of plate observations (640 fibers each) & 2564 \\
 \quad Legacy survey plates & 1802 \\
 \quad SEGUE and special plates & 676 \\
 \quad Repeat observations of plates & 86\\
 Total number of spectra\tablenotemark{b}& 1,630,960\\
 \quad  Galaxies   &  929,555 \\
 \quad  Quasars    & 121,363 \\
 \quad  Stars      &  464,261\\

 \quad  Sky        & 97,398 \\
 \quad  Unclassifiable & 28,383\\ 
 Spectra after removing skies and duplicates & 1,440,961\\
\enddata
\tablenotetext{a}{Includes regions of high stellar density, where the
  photometry is likely to be poor.  See text for details.  This area
  also includes some regions of overlap.}
\tablenotetext{b}{Spectral classifications from the {\tt spectro1d}
  code; numbers include duplicates. }
\end{deluxetable}

\begin{figure}[t]\plotone{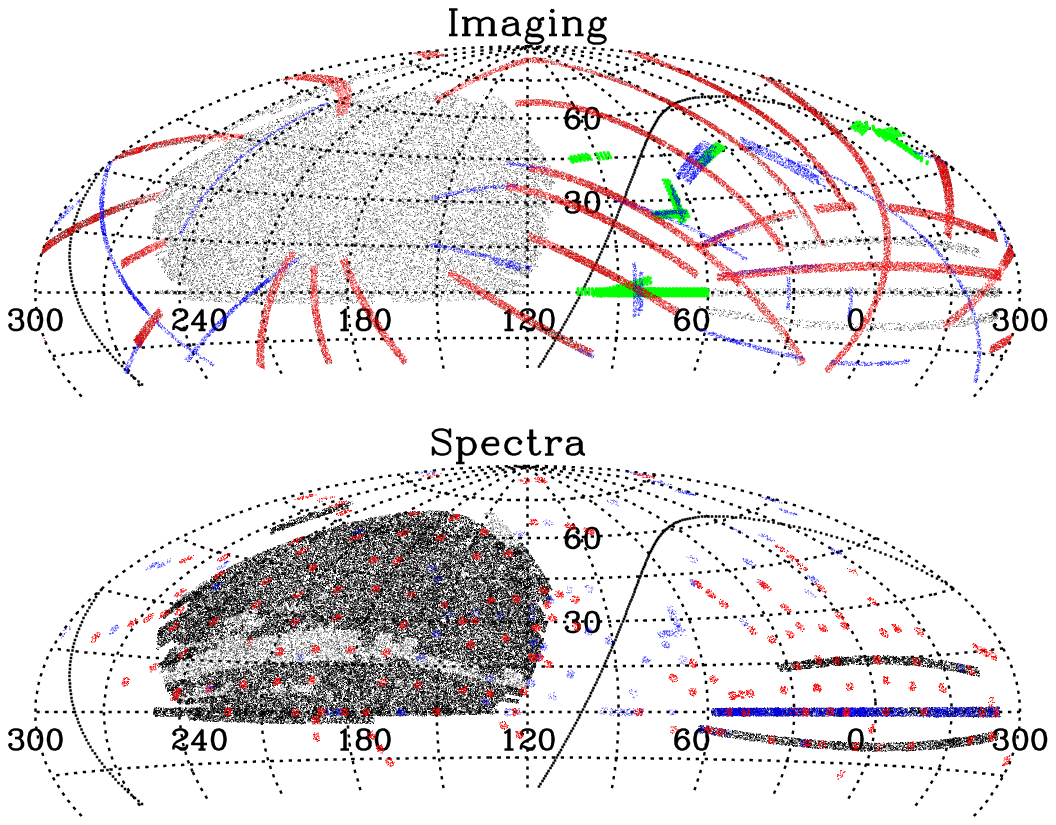}\caption{
The distribution on the sky of the data included in DR7 (upper panel:
imaging; lower panel: spectra), shown in an
Aitoff equal-area projection in J2000 Equatorial Coordinates.  The
Galactic Plane is the sinuous line that goes through each panel.  The
center of each panel is at $\alpha = 120^\circ \equiv 8^{\rm 
  h}$, and the plots cut off at $\delta = -25^\circ$, below which
the SDSS did not extend.  
The Legacy imaging survey covers the contiguous
area of the Northern Galactic Cap (centered roughly at $\alpha =
200^\circ, \delta = 30^\circ$), as well as three stripes (each of width
$2.5^\circ$) in the Southern Galactic Cap.    In addition, 
several stripes (indicated in blue in the imaging data) are auxiliary
imaging data, while the
SEGUE imaging scans are indicated in red.  The green scans
are additional runs as described in Finkbeiner \etal\ (2004).  In the
spectroscopy panel, the lighter regions indicate that area in the
Northern Galactic Cap which is
new to DR7; note that the Northern Galactic Cap is now contiguous.
Red points indicate SEGUE plates, and blue points indicate other
non-Legacy plates (mostly as described in the DR4 paper). 
\label{fig:skydist}}\end{figure}

The imaging for the Legacy survey was substantially complete with
DR6.  In DR7, we include imaging of a few small gaps that were missed
in the contiguous region of the North Galactic Cap, and repeat
observations of a few regions of the sky
which had particularly poor seeing in previous data releases.  The
total footprint has increased by 
less than 10 deg$^2$ in total.  The Legacy imaging footprint is
visible as the large contiguous gray area on the left side of the
upper panel of Figure~\ref{fig:skydist}, together with the three gray
stripes visible on the right side.  The principal augmentation of the
imaging data in DR7 are the stripes which are part of the SEGUE
survey.  They are indicated in red in the figure, and increase the
SDSS imaging footprint by roughly 2000 deg$^2$ over DR6.  Note that
many of these cross the Galactic plane (indicated by the sinuous line
crossing the figure).  Unlike DR6, the union of the Legacy and SEGUE
data are now available in a single database in CAS in DR7.  

These data
have been recalibrated using ubercalibration (Padmanabhan \etal\
2008) using the overlap between adjacent scans; the resulting
photometry is now the default photometry found in 
the CAS.  We also make available the original photometry calibrated by
the auxiliary Photometric Telescope (Tucker \etal\ 2006).  The 
ubercalibration solution was regenerated using all the imaging data,
but the changes are tiny from the ubercalibration results published in
DR6: 0.001 mag rms in 
$griz$ and 0.003 mag in $u$.  The ubercalibrated photometry zeropoints
are defined to be the same as that measured from the Photometric
Telescope.  

The green and blue patches indicate supplementary imaging stripes,
which contain scans over M31 or in its halo, through the center of the
Perseus cluster of galaxies, over the low-latitude globular cluster
M71, near the South Galactic Pole, along the orbit of the Sagittarius
Tidal Stream, and through the star-forming regions of Orion
(Finkbeiner \etal\ 2004).  In addition, there are a number of scans
at angles perpendicular, or at an oblique angle, to the regular
Legacy or SEGUE imaging stripes. These scans are used in the
ubercalibration procedure to tie the zeropoints of the stripes
together and to determine the flat-fields.

  The lower panel in Figure~\ref{fig:skydist} shows the coverage of
  spectroscopy in DR7; the light gray area shows the increment in the Legacy
  survey over DR6.  Most importantly, the gap cutting the North
  Galactic Cap in two pieces
  in previous data releases
  has been closed; we now have complete spectroscopy of our principal
  galaxy and quasar targets over a contiguous area of roughly 7500
  deg$^2$.   An additional dozen plates were
  observed to fill holes in the nominally contiguous regions in DR6.
  Adding in the three stripes in the Southern Galactic Cap, the Legacy
  spectroscopy footprint is 8032 deg$^2$, a 26\% increment over DR6. 

\begin{figure}\centering\includegraphics[width=12cm,angle=270]{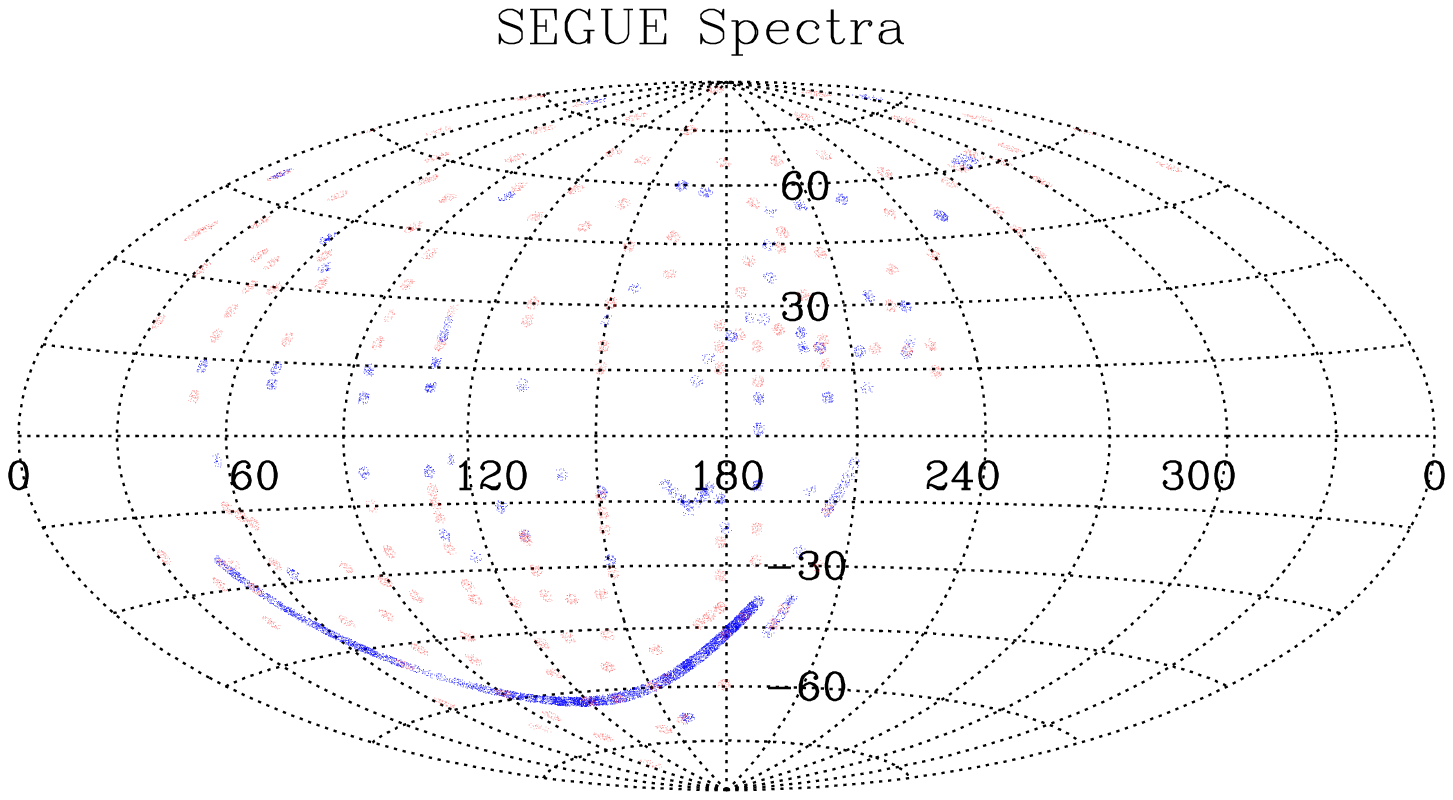}
\caption{The distribution on the sky of SEGUE (red) and other
  non-Legacy (blue) spectroscopic observations, here plotted in
  Galactic coordinates.  The contiguous blue stripe across the bottom
  is Stripe 82, along the Celestial Equator.  As described in the DR4
  paper, Stripe 82 includes extensive spectroscopy of a number of different
  types of targets outside the Legacy survey. 
\label{fig:skydist_segue}}\end{figure}

  In addition, spectroscopy was carried out using a series of target
  selection algorithms designed to find stars of a wide variety of
  types as part of the SEGUE project (DR6 paper; Yanny \etal\ 2009).
  These targets were drawn from 
  both the SEGUE and Legacy imaging, and are shown in red in the lower
  panel of Figure~\ref{fig:skydist}.  As some of these are lost in the
  density of Legacy spectra, we show the distribution of SEGUE and
  other non-Legacy spectra in Galactic coordinates in 
  Figure~\ref{fig:skydist_segue}.  

  Finally, as described in Yanny \etal\ (2009), we carried out
  spectroscopy of stars in 12 open and globular clusters to
  calibrate the measurements of stellar parameters in SEGUE (Lee
  \etal\ 2008a,b).  Many of
  these clusters are sufficiently close that the giant branches are brighter
  than the photometric saturation limit of SDSS, so the targets for
  these plates were
  selected from the literature.  Indeed, the spectrographs would
  saturate as well with our standard 15-minute exposures,
  so these observations had individual exposure times as short as
  one or two minutes. Without proper flux calibrators or exposure of
  bright sky lines to set the zeropoint of the wavelength scale, the spectrophotometry and
  wavelength calibration of the spectra on these plates are often quite
  inferior to that of the main survey, and these 
  plates are available only in the DAS, not the CAS. 

\begin{figure}\plotone{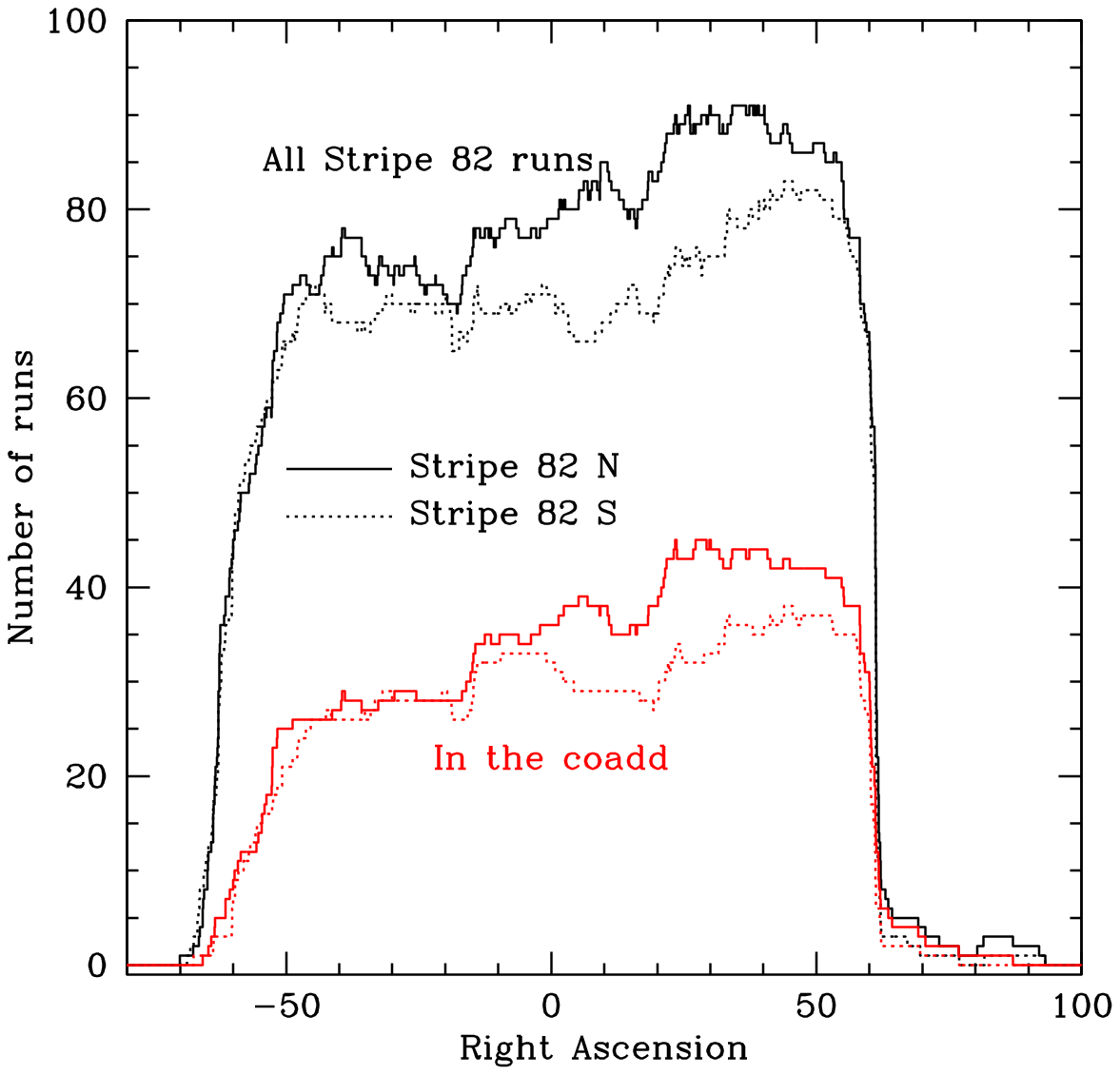}\caption{
Stripe 82, the Equatorial stripe in the South Galactic Cap, has been
imaged multiple times.  The lower pair of curves 
show the number of
scans covering a given right ascension in the North and South strip
that are included in the coaddition (mostly data taken through 2005).  
In addition, Stripe 82 has been covered many more times as part of
a comprehensive survey for $0.05 < z < 0.35$ supernovae, although often
in conditions of poor seeing, bright moon, and/or clouds; the total
numbers of scans at each right ascension in the North and South
strip are indicated in the upper pair of curves.  All these data have
been flux-calibrated, as discussed in the text, and are available
(together with the coadd itself) in the {\tt stripe82} database. 
\label{fig:stripe82}}\end{figure}

  As described in more detail below, the $2.5^\circ$ stripe centered
  on the Celestial Equator was imaged multiple times throughout SDSS
  and SDSS-II.  Each 2.5$^\circ$ wide stripe is observed by a pair of
  offset {\em strips} to cover the full width (York \etal\ 2000); the
  coverage of the two strips of Stripe 82 is shown 
  in Figure~\ref{fig:stripe82}.  The data are shown both for the
  subset of data included in a deep coaddition (lower set of curves;
  \S~\ref{sec:coadd}),
  and all scans, including those taken under non-ideal 
  conditions for the supernova survey (\S~\ref{sec:stripe82}; Frieman
  \etal\ 2008).   

\section{Additional Imaging Products and Databases}
\label{sec:runsDB}

\subsection{The Runs Database}
The SDSS imaging survey was primarily designed to give a single pass
across the sky, thus in the CAS, each photometric measurement is
flagged either {\tt Primary} or {\tt Secondary}.  {\tt Primary}
objects designate a unique set of detections (i.e., without
duplicates) using the geometric boundaries of survey
stripes\footnote{See {\tt http://www.sdss.org/dr7/algorithms/resolve.html}
  for a detailed explanation.}.  The set
of {\tt Secondary} objects includes repeat observations of the same
object in overlapping strips an stripes.  {\tt Primary} objects are
associated with a run and field which is the primary 
source of imaging data at that position.  In DR7, the union of the
Legacy and SEGUE footprints serves as the {\tt Primary} footprint; a
quantity {\tt inLegacy} in the {\tt fieldQA} table in CAS indicates
those objects which lie within the original Legacy Northern
Galactic Cap Survey ellipse, as defined in York \etal\ (2000).  
Legacy imaging can also be distinguished by the stripe number for each
run; 
stripes 9-44,
76, 82 and 86 are in the Legacy survey, all others are SEGUE stripes
or other miscellaneous pieces of sky (Figure~\ref{fig:skydist}).

While resolving the sky into a
seamless {\tt Primary} region of unique detections of objects is ideal for
many science queries, it is sometimes convenient to query
data by run without regard to the way the survey resolves overlaps and
imposes the boundaries of the edge of the survey.  These boundaries
are restricted to matched pairs of North and South strips in the main
DR7 CAS.  Therefore in many runs,  several fields at the
beginning or end which do not have a match in the corresponding
other strip are not included in the main CAS.  Thus we have
now made available a separate {\tt runs} database within the CAS,
which includes {\em all} fields in all runs, and which allows one
to query objects by which run they are imaged in.

The {\tt runs} database contains 530 complete
runs from SDSS-I and SDSS-II, where {\tt Primary} is set strictly based on
geometric limits within each scan, regardless of overlapping runs or stripes.  
The {\tt runs} database also contains 
several scans outside the regular DR7 Legacy or SEGUE
footprints.  For example, Stripe 205 is covered by runs
4334, 4516, 6751 and 6794, and follows the Sagittarius Stream, which
is in three pieces, the first running from $(\alpha,\delta) =
(240^\circ, -15^\circ)$ to 
$(200^\circ,+10^\circ)$, the second centered at
$(135^\circ,+35^\circ)$, and the last (overlapping several other runs)
which ends at $(45^\circ,+10^\circ)$. 

\subsection{The Stripe 82 Database}
\label{sec:stripe82}

  The SDSS stripe along the Celestial Equator in the Southern Galactic
  Cap (``Stripe 82'') was imaged multiple times in the Fall months.
  This was first carried out to allow the data to be stacked to reach
  fainter magnitudes, and through Fall 2004, these data were taken only under
  optimal seeing, sky brightness, and photometric conditions (i.e.,
  the conditions required for imaging in the Legacy Survey; York
  \etal\ 2000).  There
  were 84 such runs made public in previous data releases.  In Fall
  2005, 2006, and 2007, 219 additional imaging runs were taken on
  Stripe 82 as part of the SDSS supernova survey (Frieman \etal\
  2008), often under less optimal conditions: poor seeing, bright
  moonlight, and/or non-photometric conditions.  These data have been
  photometrically calibrated following the prescription of Bramich
  \etal\ (2008), whereby the photometry of bright stars is tied to that
  of photometric data on a field-by-field basis (see Ivezi\'c \etal\ 2007 for
  a similar approach).  Bramich \etal\ solved for photometric offsets
  both parallel and perpendicular to the scan direction in data from a given
  CCD; we found that the term perpendicular to the scan direction
  added little, and we did not include it here.  As Bramich \etal\
  (2008) show, the resulting photometric calibration is good to 0.02
  mag at the bright end in up to 1 mag of atmospheric extinction.  Of course,
  under non-optimal conditions, these data will not necessarily reach
  as deep as normal survey images.

SDSS judges photometricity of a given night by monitoring fluctuations
in the night sky measured by a wide area infrared camera (the ``cloud
camera'') sensitive at
10$\mu$m, where clouds are emissive (Hogg \etal\ 2001).  If the sky
fluctuations are 
small and constant, then the night is photometric.  Clouds cause the
fluctuations to increase.  Plots of cloud cover and seeing for most
nights on which Stripe 82 was observed are available as part of the
DR7 web documentation listing all Stripe 82 scans.
In addition, for those runs which the cloud camera indicated as
non-photometric, we examined the fluctuations in the zeropoint for each
CCD in the camera as a function of time using the photometric
calibration procedure of Bramich \etal\ (2008).  These zeropoint
values are available in the CAS; rms variations of more than 0.1 mag
are an indication of considerable variable cloud cover, 
and a value of more than 1 magnitude suggests that the approximate
calibration procedure of Bramich \etal\ (2008) breaks down, and the
resulting photometry should be regarded with caution. 
  All 303 runs covering Stripe 82 are made available as part of the
  {\tt Stripe82} database, which is structured like the {\tt runs}
  database.  

\subsection{Going Deep on Stripe 82}
\label{sec:coadd}

  We have carried out a coaddition of the repeat
  imaging scans on Stripe 82 taken through Fall 2005 under the best
  conditions (see below). The coaddition includes a
  total of 122 runs, covering any given piece of the $>250$ deg$^2$
  area between 20 and 40 times (Figure~\ref{fig:stripe82}), and the
  results are made available in the {\tt Stripe82} database as well.
  The coaddition runs are designated 100006 (South strip) and 
  200006 (North strip) respectively in the DAS, and 106 and 206 in the
  CAS. 

The coaddition is described in detail in Annis \etal\ (2009); see also
Jiang \etal\ (2008).
From the list of runs on 
Stripe 82 taken through the Fall 2005 season, all fields with
seeing in the $r$ band worse than $2''$ FWHM, $r$-band sky
brightness brighter than 19.5 magnitudes in one square arcsecond, or whose
photometric correction \`a la Bramich \etal\ (2008; see above) was
greater than 0.2 mags were excised; this rejected 32\% of the
available data.  The individual runs were remapped onto a uniform
astrometric coordinate system.  Interpolated pixels in each individual
run (e.g., for bad columns, bleed trails, and cosmic rays) were masked
in the coaddition process.  The sky was subtracted from each frame,
and the images coadded with weights for each frame proportional to the
transparency and inversely proportional to the square of sky noise and
seeing on each frame.  Strongly discrepant pixels were clipped in the
coaddition.  The effective seeing FWHM is 
$\sim 1.2''$ (for the southern strip of the stripe) and $\sim 1.3''$ (for the
northern strip).  

  The resulting coadded images were run through the SDSS photometric
  pipeline, yielding the catalogue made available in the {\tt Stripe82}
  database.  Rather than deriving the Point Spread Function (PSF) from
  scratch, we synthesized the PSF at each point in the sky by taking
  the suitably weighted sum of the PSFs output by the SDSS photometric
  pipeline from each of
  the individual runs.  

  Color-color diagrams of stars and counts of stars and
  galaxies as a function of magnitude demonstrates that the photometry
  reaches roughly two magnitudes fainter than single SDSS scans, similar
  to what is expected given the number of runs in the coadd.  We have
  found that star-galaxy separation is improved over that in the
  single scans, in that the cut can be made closer to the stellar
  locus.  In the main survey, objects with $m_{PSF} -
  m_{model} > 0.145$ are flagged as galaxies in a given band.
  However, the stellar peak in the PSF -- model magnitude difference
  distribution in the coadd is much narrower, allowing objects with $m_{PSF} -
  m_{model} > 0.03$ in $r$ to be flagged as galaxies. 

  The coaddition does not properly propagate information on saturated
  pixels in individual runs, and therefore the photometry of objects
  brighter than roughly $r=15.5$ is suspect.  Unfortunately, there is
  no processing flag that one can use to identify such data; we
  recommend a simple magnitude cut.  

  The SDSS photometry is quoted in terms of asinh magnitudes, as
  described by Lupton, Gunn, \& Szalay (1999), whereby the logarithmic
  magnitude scale transitions to a linear scale in flux density $f$ at low signal-to-noise
  ratio:
\begin{equation}
\label{luptitude}
m = -\frac{2.5}{\ln 10} \left[{\rm asinh}\left(\frac{f/f_0}{2\,b}\right) 
+ \ln(b)\right]. 
\end{equation}
 The magnitude at which this transition occurs is set by the quantity
 $b$, which is roughly the fractional noise in the sky in a PSF
 aperture in $1^{\prime\prime}$ seeing (EDR paper).  Here $f_0 = 3631$
 Jy, the zeropoint of the AB flux scale.  The quantity $b$ for the
 coaddition is given in  Table~\ref{table:asinh}, along with the asinh
 magnitude associated  with a zero flux object.  Compare with the
 equivalent numbers for the main survey, given in Table 21 of the EDR
 paper.  Table~\ref{table:asinh} also
lists the flux corresponding to $10\,f_0 b$, above which the asinh
magnitude and the traditional logarithmic magnitude differ by less
than 1\% in flux.  

\begin{deluxetable}{cccc}
\tablecaption{Asinh Magnitude Softening Parameters for the Coaddition\label{table:asinh}}
\tablecolumns{4}
\tablehead{
  \colhead{Band} & \colhead{$b$}  
& \colhead{Zero-flux magnitude}
	& \colhead{$m$}\\
&&($m(f/f_0=0)$)&$(f/f_0=10b)$}
\startdata
$u$ & $1.0 \times 10^{-11}$ & 27.50 & 24.99\\
$g$ & $0.43 \times 10^{-11}$ & 28.42 & 25.91\\
$r$ & $0.81 \times 10^{-11}$ & 27.72 & 25.22\\
$i$ & $1.4 \times 10^{-11}$ & 27.13 & 24.62\\
$z$ & $3.7 \times 10^{-11}$ & 26.08 & 23.57
\enddata
\end{deluxetable}

  As with the main survey, it is important to use the various
  processing flags output by the photometric pipeline (e.g., as
  recommended by Richards \etal\ 2002) to reject
  spurious objects, and to select
  objects with reliable photometry.

\section{Improvements in Processing of Imaging data}
\label{sec:imaging}

\subsection{New Reductions of SEGUE Imaging Data and Crowded Fields}

As was noted in the DR6 paper, the SDSS imaging pipeline (\photo) was
designed to analyze data at high Galactic latitudes, and is not
optimized to handle very crowded fields.  The Legacy survey is
restricted to high latitudes, and \photo\ performs adequately
throughout the Legacy footprint. However, at lower latitudes, when the
density of stars brighter than $r=21$ grows above 5000 deg$^{-2}$, the
pipeline is known to fail, as it is unable to
find sufficiently isolated stars to measure an accurate PSF, and
the deblender does poorly with overly crowded images.  Many of the
SEGUE scans probe these low 
latitudes (Figure~\ref{fig:skydist}), and we therefore adapted an
alternative stellar photometry code called \ps\ developed by the
Pan-STARRS team (Kaiser \etal\ 2002; Magnier 2006) to be used for
these runs.  In brief, we first run this code, and then run \photo\ 
using the list of objects detected by \ps\ as input to help \photo's
object finder in crowded regions.  This approach thus provides two 
sets of photometry at low latitudes. 

Like e.g., DAOPHOT (Stetson 1987), \ps\ begins with the assumption
that every object is unresolved, and therefore does a better job than
\photo\ in crowded stellar regions.  It uses an analytical model based
	   on Gaussians 
to
describe the basic PSF shape, with parameters which may vary across
the field of the image to follow the PSF variations.  
It also uses a pixel-based 
representation of the residuals between the PSF objects and the
analytical model, which is also allowed to vary across each field. 
Candidate PSF stars are selected from the collection of bright objects
in the frame by searching for a tight clump in the distribution of
second moments.  After rejecting outliers, the PSF fit parameters are
used to constrain the spatial variations in the PSF model.

Unlike \photo, \ps\ processes each frame separately
(without any requirement of continuity of PSF estimation across frame
boundaries), and each filter separately (without any requirement that
the list of objects between the separate filters agree).  The pipeline
outputs positions and PSF magnitudes (and errors) for each detected
object; the results are found in the {\tt PsObjAll} table in the CAS.
The resulting photometry is then matched between filters 
using a $1^{\prime\prime}$ matching radius.   While the estimated PSF
errors output by \photo\  include a term from the uncertainty in
the PSF fitting, this component is not included in the errors
reported by \ps. 

 We then run \photo, asking it to carry out photometry at the position
 of each object detected by \ps, in addition to the positions of
 objects \photo\ itself detects.  This allows \photo\ to do a much better
 job of distinguishing individual objects in crowded regions.
 In addition, the pipeline is fine-tuned to less aggressively look for
 overlap between adjacent objects, and not to give up as soon as it
 does at high latitude when faced with deblending large numbers of
 objects.  We describe below how the photometry directly out of \ps,
 and that from \photo, compare. 

  The SDSS PSF photometry had an offset applied to it to make it agree
  with aperture photometry of bright stars within a radius of
  $7.43^{\prime\prime}$; this large aperture photometry was in fact
  what was used by ubercalibration to tie all the photometry
  together (Padmanabhan \etal\ 2008).  In crowded regions, finding
  sufficiently isolated stars to measure aperture photometry becomes
  difficult.  \ps\ photometry was forced
  to agree with these large-aperture magnitudes for bright stars; this
  was done in practice by determining, for each CCD in the imaging
  camera for each run, the
  average aperture correction needed to put the two on the
  same system, using stars at Galactic latitude $|b| > 15^\circ$, where
  crowding effects are less severe. 

If any part of a SEGUE imaging run extended to $|b| < 25^\circ$, the
entire run was processed through the \photo\ and \ps\ code.  This
sample includes most (but not all) of the SEGUE imaging runs.
These \ps+\photo\ processed runs, designated with rerun=648 in the DR7
CAS and DAS, are declared the {\tt Best} reduction of these SEGUE
runs.  There is also an inferior {\tt Target} version of these SEGUE
runs which was used to design SEGUE spectroscopic plates; it is based
on \photo\ alone, as the \ps\ code was unavailable at the time the
plates were designed.  The {\tt Target} reductions have rerun=40, and
are segregated to the {\tt SEGUETARGDR7} database.

  This processing revealed a problem with \photo.  In
  crowded regions, one cannot find sufficiently isolated stars to
  measure counts through such a large aperture, and in practice, the
  code corrected PSF magnitudes to an aperture photometry  radius of
  $3.00^{\prime\prime}$ 
  instead, whenever any part of a given run dipped below $|b| =
  8^\circ$.  Thus the aperture correction was underestimated,
  typically by 0.03--0.06 mag, depending on the seeing.  This was not a
  problem for any of the Legacy  
  imaging scans, but is very much an issue for the SEGUE runs.
  Fortunately, there is a strong correlation, in a given detector, between the
  aperture correction from a $3.00^{\prime\prime}$ aperture to a
  $7.43^{\prime\prime}$ aperture (as measured on
  high-latitude fields), and the seeing. We
  therefore applied this correction after the fact to the \photo\ 
  PSF, de Vaucouleurs, exponential, and model magnitudes for all SEGUE
  runs affected by this 
  problem.  This was carried out before ubercalibration, so these
  runs are photometrically calibrated in a consistent way.  

\subsection{Comparison of \photo\ and \ps\ Photometry}
\label{sec:sdssps}

\newcommand{\Egri}{\ensuremath{{E}_{{gri}}}}
\newcommand{\Eriz}{\ensuremath{{E}_{{riz}}}}
\newcommand{\Qgri}{\ensuremath{{Q}_{{gri}}}}
\newcommand{\Qriz}{\ensuremath{{Q}_{{riz}}}}

The quality of the photometry produced by \ps\, and by \photo\ with
the \ps-detected objects as input, 
was evaluated by comparing the
magnitudes computed by the two methods.  Within each field, we calculated the
median of the difference of PSF magnitudes for stars with $14 <
u,g,r,i,z < 20$.  This median difference had an rms of 0.014
mag.  Fields with a difference greater than 0.08 mag are suspect, and
further investigation is needed to determine which of the two
pipelines might be at fault. We followed McGehee \etal\ (2005) to
measure reddening-free colors of the same stars that track the stellar
locus: 
\begin{eqnarray}
  \label{eq:redfree}
  \Qgri &=& (g - r) - \Egri\, (r - i),  \\\nonumber
  \Qriz &=& (r - i) - \Eriz\, (i - z),  \\
\end{eqnarray}
where $\Egri = 1.582$ and $\Eriz = 0.987$. 
These are normalized to equal zero at high Galactic latitude (note
that these colors do {\em not} include the $u$ band).  
    
\begin{figure}[t]\centering\includegraphics[width=12cm]{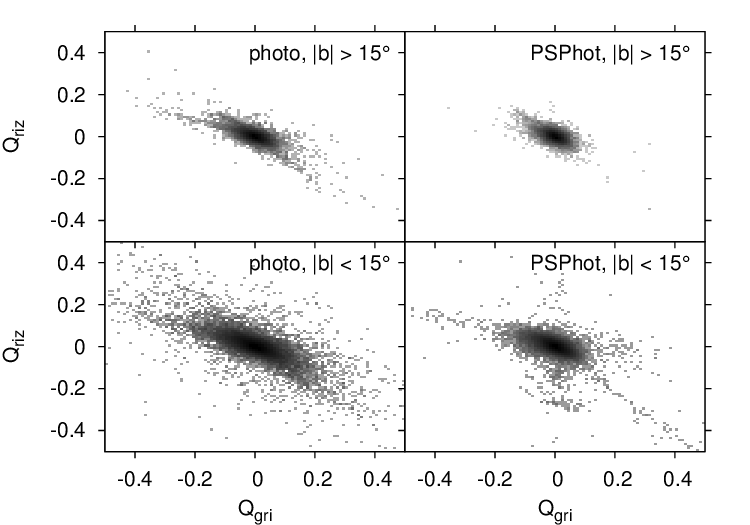}
  \caption{The distribution of median \Qgri\ and \Qriz\ parameters
    measuring the position of the stellar locus within each field for
    the \photo\ (left) and \ps\ (right) photometric pipelines; zero
    values are indicative of uniform photometry.  
    Within the Galactic plane (lower panels), the \ps\ values
    are more concentrated, but contain a higher number of systematic
    departures from the main locus.  The \ps\ code in fact gives a
    tighter locus at high latitudes as well (upper panels).  Histogram equalization
    of the gray-scale was used to emphasize low density regions. 
\label{fig:Qparams}}
\end{figure}

Median \Qgri\ and \Qriz\ colors in each field were computed for objects identified as
stars in each field, and satisfying magnitude and color cuts as follows:
$14<(u,g,r,i,z)<20$, $0.5<(u-g)<1.9$,
$0.0<(g-r)<1.2$, $-0.2<(r-i)<0.8$, and $-0.2<(i-z)<0.6$.  The
$Q$-parameters were found to be lower by up to 0.1 mag at low Galactic
latitudes; to remove this effect, we fit a model of a
constant plus Lorentzian to the median $Q$ values as a function of
Galactic latitude, and subtracted it.  
The distributions of the \Qgri\ and \Qriz\ values for both \photo\ and \ps\
are compared as density plots in Figure~\ref{fig:Qparams}. From
equation~\ref{eq:redfree}, photometric errors in a single filter
manifest themselves differently: $\delta g$ as a shift in \Qgri,
$\delta r$ as a line with slope $d\Qriz/d\Qgri = -1/(1+\Egri) = -0.35$,
$\delta i$ as a line with slope $d\Qriz/d\Qgri = -(1+\Eriz)/\Egri =
-1.07$, and $\delta z$ as a shift in \Qriz.

The \photo\ data in a given
field was flagged as bad when either $|\Qgri|$ or  $|\Qriz| > 0.12$ 
mag ($\gtrsim5\sigma$) as measured from \photo\ magnitudes, and
similarly for the \ps\ outputs.  Of course, a field could be flagged
as bad in both sets of outputs.  By this
criterion, about 2\% of the fields processed with \ps\ were flagged
bad based on the \photo\ outputs, and 
3.6\% were bad based on \ps\ photometry.  The vast majority of the flagged
fields are within $15^\circ$ of the Galactic plane, and essentially
all the fields in which the median difference between \photo\ and \ps\
photometry was greater than 0.08 mag in a given band were flagged as
bad by the $Q$ criteria.  This flag and the \Qgri\ and \Qriz\
quantities themselves can be found in the {\tt fieldQA} table in the
CAS. 

Although more fields are flagged based on the \ps\ outputs, the \ps\
scatter in Figure~\ref{fig:Qparams} is tighter at both high and low
Galactic latitudes than for \photo.  The \ps\ stellar photometry is
therefore preferred for studies of the stellar locus (we have not fully
assessed its robustness to outliers), but comes with the caveat that
fields flagged bad should be identified in the {\tt fieldQA} table and
be culled.

An alternative check of SDSS photometry in dense stellar fields was carried
out by An \etal\ (2008), who reduced the SDSS imaging
data for crowded open and globular cluster fields using the
DAOPHOT/ALLFRAME suite of programs (Stetson 1987, 1994).  
At a stellar density of $\sim400$ stars deg$^{-2}$ with $r<20$, they found
$\sim2\%$ rms variations in the difference between \photo\ and
DAOPHOT magnitudes in the scanning direction in all five bandpasses
(see their Fig.~3).  The systematic structures 
are likely due to imperfect modeling of the PSFs in \photo, given that DAOPHOT
magnitudes exhibit no such large variations with respect to aperture photometry.
In other words, the PSF variations were too rapid for the \photo\ pipeline
to follow over a time scale covered approximately by one field ($\approx10\arcmin$
or $\approx54$~sec in time).

An \etal\ (2008) further examined the accuracy of \photo\ magnitudes in
semi-crowded fields using three open clusters in their sample.
Stellar densities in these fields were as much as $\sim10$ times
higher than those in the high Galactic latitude
fields, but \photo\ recovered $\sim80-90$\% of stellar objects
in the DAOPHOT/ALLFRAME catalog.  An \etal\ (2008) found that these fields have
only marginally stronger spatial variations in \photo\ magnitudes than
those at lower stellar densities.  

\subsection{Further Assessments of Imaging Quality}
  Section 4.6 of the EDR paper describes a series of flags available
  in the database to assess the quality of each field in the imaging
  data; this includes information on whether the data in a given field
  meet survey requirements on seeing and sky brightness.  We have added
  additional criteria to assess the quality of each field. The CAS
  table called {\tt fieldQA} includes a flag called {\tt ProblemChar}
  associated with each field, which is set when:
\begin{itemize} 
  \item The median of the telescope focus over three frames moved more
    than 60$\mu$m, indicating a problem with the automated focus of
    the telescope (Gunn \etal\ 2006) ({\tt Problemchar = 'f'}); 
   \item The rotator angle moved more than $25^{\prime\prime}$ between
     adjacent fields
(corresponding to a $0.55^{\prime\prime}$ image shift at the edge of the camera)
	({\tt ProblemChar='r'});
\item The astrometric solution shifted by more than 4 pixels
  ($1.6^{\prime\prime}$) from a smooth interpolation between adjacent
  fields ({\tt ProblemChar='a'}).
\item Miscellaneous other problems, including voltage problems in the
  camera, and lights left on in the telescope; this was triggered in only
  two imaging runs  ({\tt ProblemChar='s'}). 
\end{itemize}

We flag all fields with these problems in
any of the five bandpasses. Because the imaging observations are 
done in driftscan mode (Gunn \etal\ 1998), different areas of the sky are observed
simultaneously in each bandpass and referenced to the field number of
the $r$-band observation.    Thus in the case of focus problems, we
mark the 11 fields 
preceding, and the three fields following the field in question in
all camera columns in the run as
bad.  For the rotator and astrometric shift problems, we similarly
mark the nine preceding and the one following field as bad. Only
0.3\% of all fields in the CAS are marked with one of these problems
(the majority of which are due to focus problems);
these flags should be consulted when examining the reliability of the
photometry in a given area of sky.  

\subsection{Astrometric Recalibration}
\label{sec:astrometry}

Early SDSS imaging runs were astrometrically calibrated against Tycho-2
(H\o g \etal\ 2000), which yielded statistical errors per coordinate for bright stars
($r < 20$) of approximately 75 mas and systematic errors
of 20 -- 30 mas.  Later runs were calibrated against preliminary versions
of the USNO CCD Astrograph Catalog (UCAC, Zacharias \etal\ 2000), which yielded
improved statistical errors per coordinate of approximately 45 mas,
with systematic errors of 20 -- 30 mas (Pier \etal\ 2003).  Proper
motions were not 
available for the preliminary versions of UCAC.  
Since the typical epoch difference between the SDSS and
UCAC observations is a few years and the typical proper motion of UCAC
stars is a few mas year$^{-1}$, this introduces an additional roughly
10 mas of systematic error in the positions due to the uncorrected proper
motions of the calibrating stars.  

All of DR7 has been recalibrated astrometrically against the second data
release of UCAC (UCAC2; Zacharias \etal\ 2004). While the systematic errors for UCAC2 are not yet
well characterized, they are thought to be less than 20 mas (N. Zacharias,
private communication).  UCAC2 also includes proper motions for stars with
$\delta < +41^\circ$.  For stars at higher declination, proper
motions from the SDSS+USNO-B catalog (Munn \etal\ 2004) have been merged with the
UCAC2 positions.  
With these improvements, all DR7 astrometry has
statistical errors per coordinate for bright stars of approximately 45 mas,
with systematic errors of less than 20 mas. The mean differences
per run between the old and new calibrations is a function of position on the
sky, with typical absolute mean differences of 0 to 40 mas.  The rms
differences are of order 10 to 40 mas for runs previously reduced
against UCAC, and 40 to 80 mas for runs previously reduced against
Tycho-2, consistent with what we would expect given the errors in the
reductions.  

Note that the formal SDSS names of objects in the CAS are of the form
SDSS$\,$Jhhmmss.ss$\pm$ddmmss.s.  Because of the subtle changes in the
astrometry, these names will be slightly different for many objects
between DR6 and DR7.  The user should be aware of this in comparing
objects between DR6 and DR7. 

The CAS includes proper motions for objects derived by combining SDSS
astrometry with USNO-B positions, recalibrated against SDSS (Munn
\etal\ 2004).  These are given in the {\tt ProperMotions} table in the
CAS\footnote{This table was called {\tt USNOB} in the DR3 and DR4 versions of the
CAS.}.  An error was discovered in the proper motion code in Data Releases
3 through 6, which causes smoothly varying systematic errors, in the
proper motion in right ascension only, of typically 1---2 mas
year$^{-1}$ (see Munn \etal\ 2007 for a full description of the problem
and its effects).  This error has been corrected in DR7, thus any use of
proper motions should use the DR7 CAS.

\subsection{SEGUE Target Selection}

Several of the SEGUE target selection algorithms evolved during the course 
of SDSS-II.  The most significant changes occurred to the K-giant 
algorithm, as it was realized that good color-based luminosity separation 
could be done only for the very reddest ($g-r > 1.1$) giant candidates
by their deviation from the main sequence locus in the $ugr$ color
diagram; this of course requires accurate $u$ band photometry. 
Early K giant target selection included stars with $(g-r)_0$ (where
the subscript refers to values after correcting for SFD Galactic
extinction) as blue as 0.35.  The final selection
chose stars with $(g-r)_0$ between 0.5 and 1.2, and was restricted to
$g_0 < 18.5$; this gives much cleaner samples of K giants (Yanny
\etal\ 2009).

In order to allow users to analyze completeness and efficiency of
SEGUE stellar target selection samples, the latest (v4.6) version of
the algorithms (Yanny \etal\ 2009) was applied to {\em all} stellar
objects in the imaging catalog which had $g < 21$ or $z < 21$, over
the entire sky.  
The appropriate bits were propagated into the {\tt SEGUEPrimTarget} and
{\tt SEGUESecTarget} fields of the {\tt photoObjAll} table of the DR7 CAS.
A description of the bits and the target selection algorithms is
available in Yanny \etal\ (2009). 

\subsection{Photometric Redshifts}

As described in the DR5 paper, the SDSS makes available the results of
two different photometric redshift determinations for galaxies, one based
on neural nets and the other based on a template-fitting approach. 
With DR7, we include improvements to both, as we now describe.  

\subsubsection{Photometric Redshifts with Neural Nets}

The neural net solutions for photometric redshifts and their errors (listed as {\tt
  Photoz2} in the CAS, and described in Oyaizu \etal\ 2008) have not
changed since DR6, and do not use the ubercalibrated magnitudes.
However, we now provide a value-added
catalog containing the redshift probability distribution for each  
galaxy, $p(z)$, calculated using the weights method presented in
Cunha \etal\ (2008).  The $p(z)$ for each galaxy in the catalog is the
weighted distribution of the spectroscopic redshifts of the $100$
nearest training-set galaxies in the space of dereddened model
colors and $r$ magnitude.  For the $p(z)$ calculation
we also added the zCOSMOS (Lilly \etal\ 2007) and DEEP2-EGS (Davis
\etal\ 2007)
galaxies to the spectroscopic training set used for the {\tt Photoz2}
solution.  

Cunha \etal\ (2008) showed that summing the $p(z)$ for a sample of
galaxies yields a better estimation of their true redshift
distribution than that of the individual photometric redshifts.  
Mandelbaum \etal\ (2008) found that this gives significantly smaller
photometric lensing calibration bias than the use of a single photometric redshift
estimate for each galaxy.  

\subsubsection{Photometric Redshifts: A New Hybrid Technique}

With DR7, we have made substantial improvements
in the other photometric redshift code ({\tt Photoz}), using a hybrid
method that combines the template 
fitting approach of Csabai \etal\ (2003; i.e., the approach used in DR5
and DR6) and an empirical calibration
using objects with both observed colors and spectroscopic redshifts.  We
summarize the method briefly here, with details to follow in a paper
in preparation. 

 The
spectroscopic sample of SDSS contains over 900,000 spectroscopically
confirmed galaxies, and the combination of the main sample (Strauss
\etal\ 2001), the LRG sample (Eisenstein \etal\ 2001) and special
plates targeted at fainter blue galaxies (DR4 paper) more or less
cover the whole 
color region in which galaxies lie to the depths of SDSS.  Thus we use
the DR7 spectroscopic set as a reference set for redshift estimation without
any additional data from synthetic spectra. 

The estimation method uses a k-d tree (following Csabai \etal\
2007) to search in the {ubercalibrated} $u-g$, $g-r$,
$r-i$, $i-z$ color space for the 100 nearest neighbors of every object
in the estimation set (i.e. the galaxies for which we want to estimate
redshift) and then estimates redshift by fitting a local hyperplane to
these points, after rejecting outliers.  
If an object lies outside the bounding box of the 100 nearest
neighbors in color space, the photometric redshift is less reliable,
and the object is flagged accordingly.

We use template fitting to estimate the K-correction, 
distance modulus, 
absolute magnitudes,
rest frame colors, and spectral type.  
We search for the best match of the
measured colors 
and the synthetic colors calculated from  repaired
(Budavari \etal\ 2000) empirical template spectra at the redshift given
by the local nearest neighbor fit. 

We have found that the mean deviations of the redshifts from the
best-fit hyperplane is a good estimate of the error.
That, together with the flag indicating whether an object lies outside
the bounding box of its neighbors,
and the difference between the estimated photometric redshift and the
average redshift of its neighbors, 
can be used to select objects 
with reliable photometric redshift values. 

\begin{figure}[t]\centering
\includegraphics[width=12cm,angle=270]{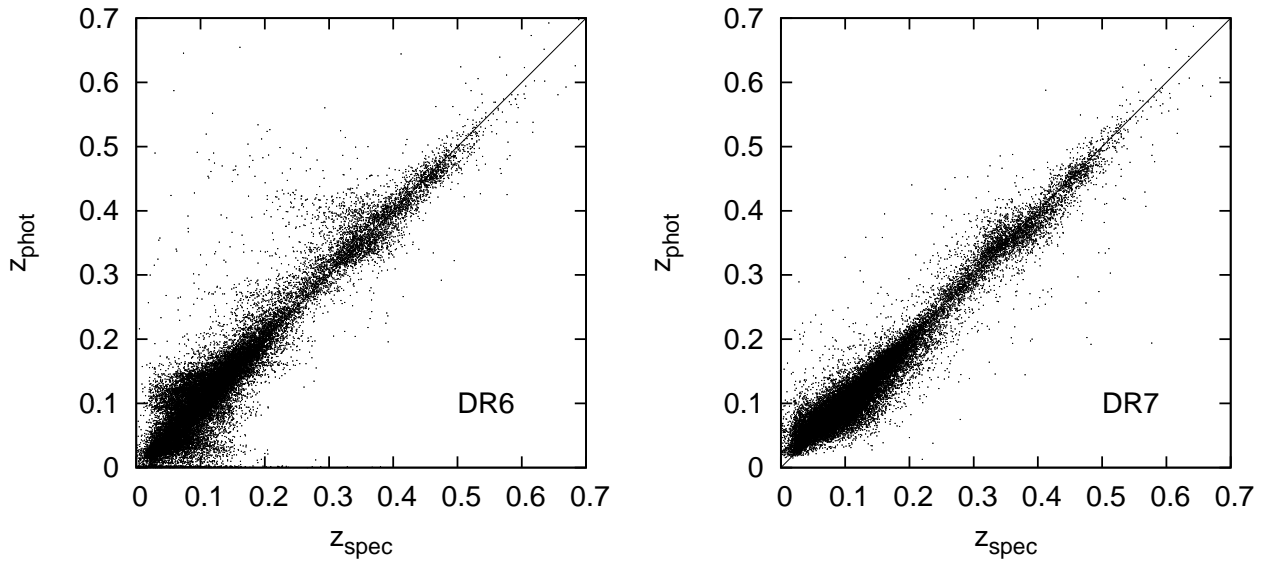}
\caption{The template-based estimated redshifts versus the true spectroscopic redshifts for a
random sample of 30,000 galaxies with redshifts from SDSS. The
estimated values calculated with the old (DR6) method has
significantly larger scatter and more outliers than the ones with the
new hybrid (DR7) technique. Note that the sample is dominated by red
galaxies (whose photometric redshifts are intrinsically easier to
measure) at $z > 0.2$. 
\label{fig:photoz}}\end{figure}

The rms error of the redshift estimation for the reference set
decreases from 0.044 in DR6 to 0.025 in DR7
with this improved algorithm (Figure~\ref{fig:photoz}).  Iteratively
removing the outliers beyond 
$3\,\sigma$ gives rms errors of 0.028 and 0.020 for
the old and new methods, respectively. In addition, the reliability of
the quoted 
errors is much higher. 

\subsection{SDSS Filter Response Functions}

The response functions of the SDSS imager as a function of wavelength
have been monitored throughout the survey.  The $griz$ responses were
stable over time, although very small seasonal (i.e., temperature)
variations were observed, at a level well below our typical
photometric errors.  However, we have found a relatively large change
in both the amplitude and shape of the $u$-band response, which is
likely due to a degradation of the UV enhanced coating of the $u$-band
CCD.  
This change in instrumental zero-point is effectively corrected by the 
photometric calibration for objects near the mean color of the standard 
stars, and, in fact, the repeat photometry of stars in stripe 82 is stable 
with time for stars with $-0.5<g-r<1.5$ (Bramich \etal\ 2008; Ivezi\'c
\etal\ 2007).  However, the observed  
response changes involve a roughly 30 \AA\ redward shift in the effective 
wavelengths of the $u$ filters over the lifetime of the survey, so one 
would expect significant changes in the measured colors of objects of 
extreme color over the period, and this is being investigated. Doi \etal\ 
(in preparation) will summarize the filter characteristics in full, 
including column-to-column variations within the camera and the changes 
with time.

\section{Photometry of Bright Galaxies}
\label{sec:skysub}
As described in the DR6 paper and Mandelbaum \etal\ (2005), systematic
errors in the estimation of the sky near bright ($r<16$) galaxies
causes their fluxes and scale sizes to be 
underestimated and the number of neighboring objects to be 
suppressed.  Indeed, a number of authors (Lauer \etal\ 2007, Bernardi
\etal\ 2007, Lisker \etal\ 2007) have pointed out systematic errors in
SDSS galaxy photometry at the bright end. In the 
DR6 paper, this effect was quantified by adding simulated galaxies to
the SDSS images using a code described in Masjedi \etal\ (2006). These
simulations found that the $r$ band brightness of galaxies was
underestimated by as much as 0.8 magnitudes for a 12th magnitude
galaxy with S\'ersic index, $n=1$ (an ``exponential'', or disk
galaxy). For $n=4$ galaxies (``de Vaucouleurs'', or elliptical
galaxies), the effect was less pronounced, with a brightness
underestimate of less than 0.6 magnitudes. 

\begin{figure}
\centering
        \includegraphics[angle=270, scale=.6]{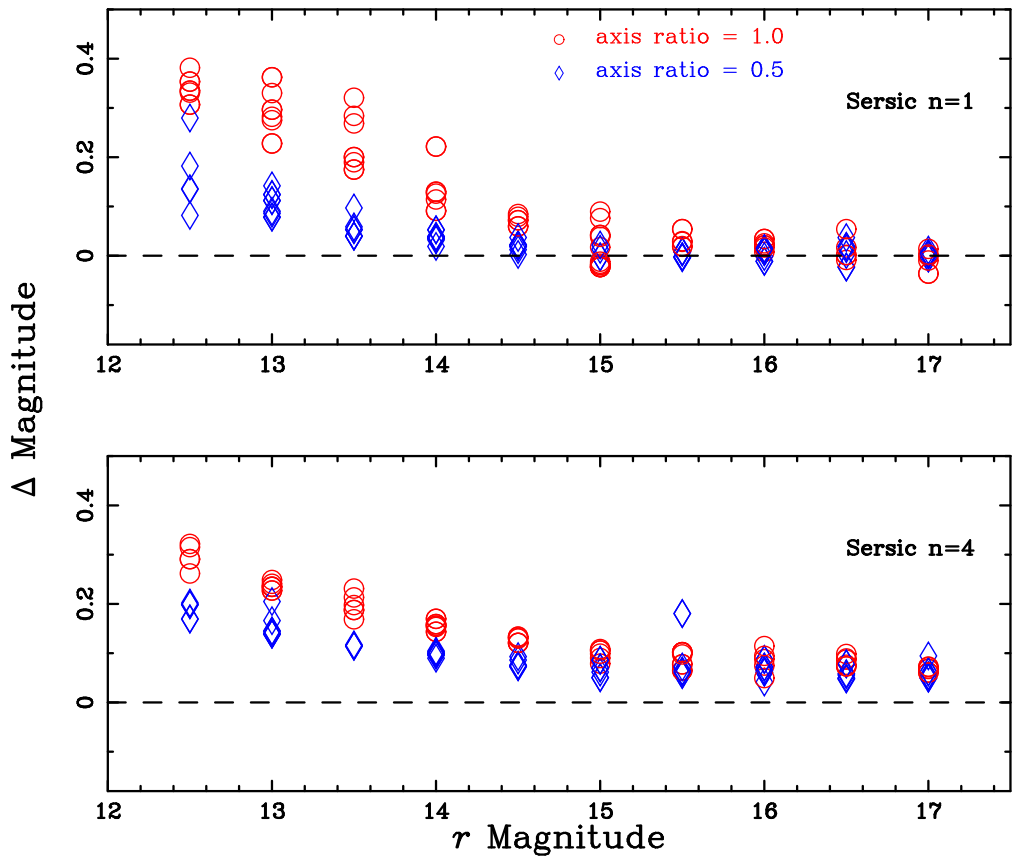}
        \caption{Difference between measured model and true $r$-band magnitudes of a series 
of simulated galaxies with S\'ersic index of 1 (disk galaxies; upper
panel) and 4 (elliptical galaxies; lower panel).  These galaxies 
followed the magnitude-effective radius relation observed in the SDSS
Value-Added Galaxy Catalog (Blanton \etal\ 2005),
and were either circularly symmetric (circles) or had an axis ratio of
0.5 (diamonds).  They were added to random areas of real high-latitude
fields, and run through \photo.  The simulated elliptical galaxies
show a systematic offset even at the faint end; this is due to the
fact that the \photo\ model magnitude code assumes a truncation beyond
7 scale-lengths, while the ``true'' magnitude has no such
truncation.  This is a 0.05 mag effect.  
\label{fig:delta_bright}}
 \end{figure}

However, the simulations shown in the DR6 paper used an incorrect
relation between galaxy size and magnitude, in the sense that they 
overestimated the extent of the problem for the typical galaxy. Using
instead the relationships between apparent magnitude and half-light
radius measured for SDSS bulge and disk galaxies (Blanton \etal\ 2003),
we repeated the
exercise: we simulated pure $n=1$ and $n=4$ galaxies with axis
ratios $b/a$ of 0.5 and 1, and added them to real $r$-band SDSS
images.  We ran the results through \photo\ and compared their
measured model magnitudes to their true magnitudes; the bias in
the measurement is shown as a function of true magnitudes in
Figure~\ref{fig:delta_bright}.  There is appreciable scatter at a
given magnitude, due both to the changing background and the different
axis ratios.  
On average, however, the flux is
underestimated by approximately 0.2 magnitudes at $r=12.5$ and ${}<0.1$
magnitudes at $r=15$ for simulated galaxies with a S\'ersic index of
1. For a S\'ersic index of 4, the flux is underestimated by as much as 0.3 
magnitudes at $r=12.5$. The effect is more severe for simulated objects with 
an axis ratio of 1 than for an axis ratio of 0.5  (see 
Figure~\ref{fig:delta_bright}).  The scale sizes of galaxies are
similarly underestimated by as much as 20\%
for simulated galaxies with S\'{e}rsic index of 1, and 30\% for
an index of 4. Of course, the most massive
elliptical or cD galaxies will have more extended envelopes, producing a
larger effect than we have found here (Lauer \etal\ 2007).

\section{Improvements in Processing of Spectroscopic Data}
\label{sec:spectroscopy}

\subsection{Correction of Instability in the Spectroscopic Flats}

Spectroscopic flatfields for the blue camera in the first spectrograph
contain an interference pattern produced by the dichroic.  The
thickness of the dichroic coating is believed to be sensitive to the
ambient humidity, and moisture which enters the system during plate
changes affects the instrument response, shifting the interference
pattern in wavelength in unpredictable ways on timescales comparable
to the 900s exposure time.  The flats applied in processing were
exposed several minutes prior to, or after, the science frames and
therefore were not always representative of the true instrument
response at the time of exposure. The interference pattern is most
pronounced in the 3800-4100\AA\ region of the spectrum.  If it
shifts during an exposure, it will not be properly corrected by the
flatfield, causing significant distortion of blue
absorption lines in stellar spectra, and systematically affecting estimates
of metallicities and surface temperatures.

\begin{figure}\centering\includegraphics[width=12cm]{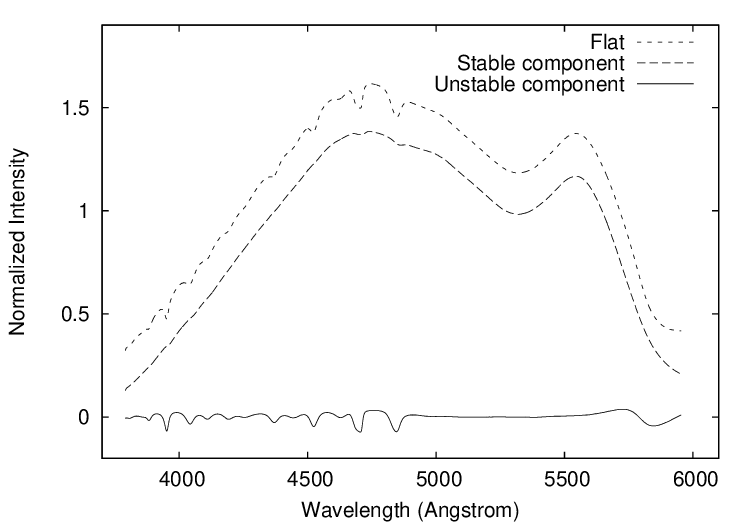}\caption
{The decomposition of the flat field of the first blue
  spectrograph (upper curve) into stable (lower curve, offset slightly
  for clarity) and unstable (interference) components. The unstable
  component is close to zero, but shows wiggles at wavelengths that
  shift from one exposure to another. 
\label{fig:superflat}}\end{figure}

\begin{figure}
\centering
\includegraphics[width=12cm]{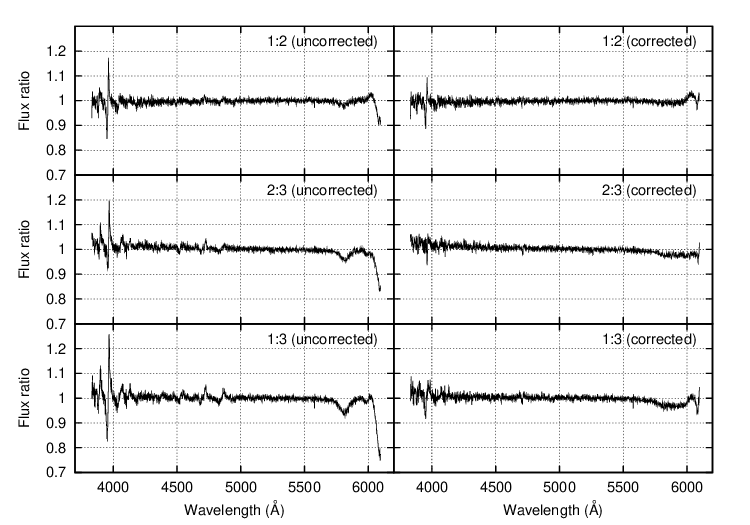}
\caption{Median flux ratios over all objects in the three exposures of
  plate 1916, before (left) and after (right) correction for the
  moving interference filters.  The ratio is fit to the derivative of
  the interference component of the flat field
  (Figure~\ref{fig:superflat}) after allowing for an arbitrary
  wavelength shift.
\label{fig:ratios}}
\end{figure}

\begin{figure}\centering\includegraphics[width=12cm,angle=270]{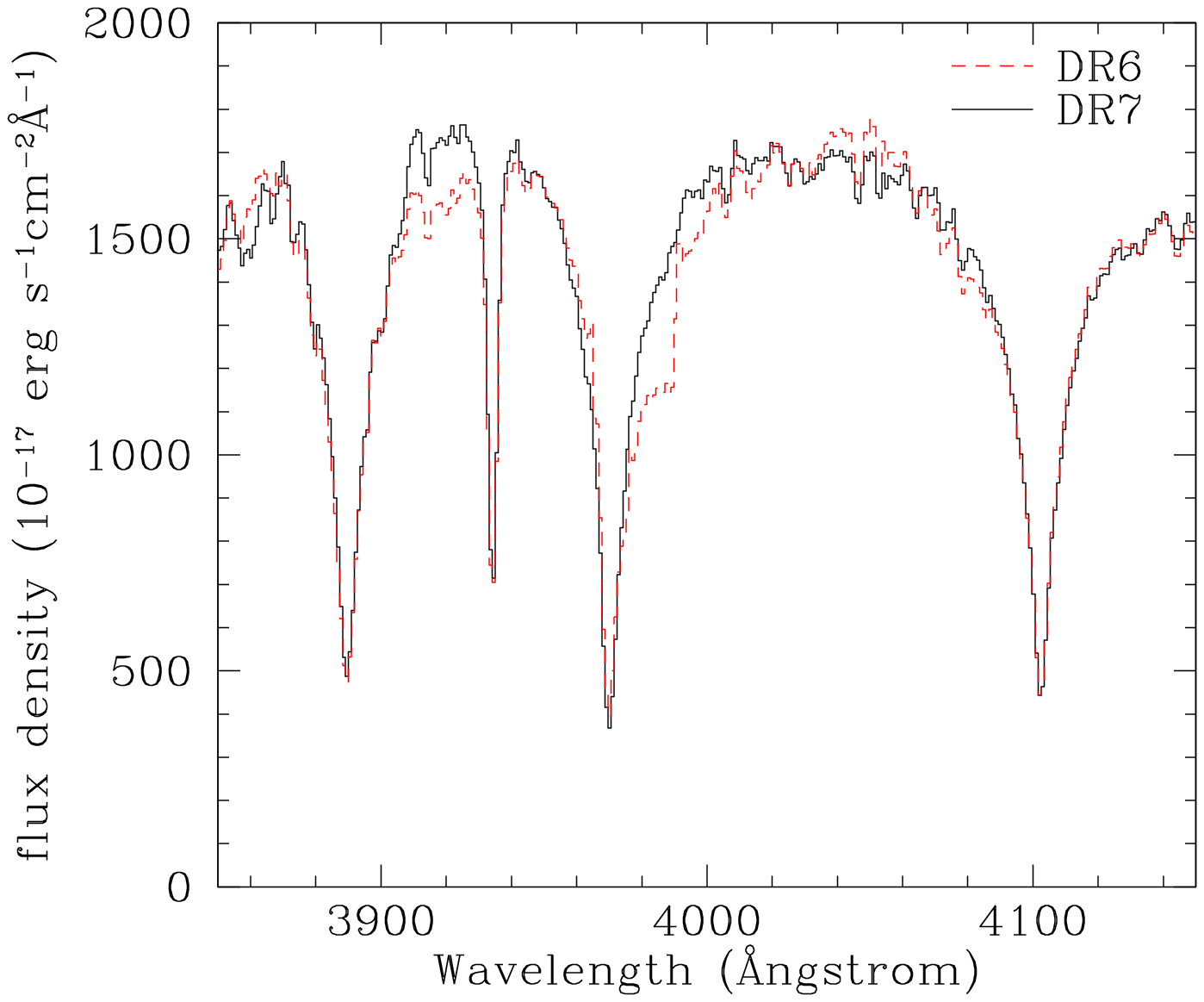}\caption{
The spectrum of SDSS$\,$J172637.26+264127.6, an A0 star observed as
part of SEGUE.  The strong broad lines are due to Balmer absorption.
The red dashed spectrum is that available in DR6, while the black
solid spectrum is from DR7, with its improved flat-field.  
\label{fig:wiggle_spectrum}}\end{figure}

Flats obtained under different conditions were used to identify and
model the stable and unstable (shifting) components of the flat, as
shown in Figure~\ref{fig:superflat}.  With this model in hand, we
searched for shifts in the interference pattern over the typically 45 
minute time a given plate was observed by comparing the results of the
individual 15-minute exposures for each object.  Thus we took ratios
of the extracted spectra from the separate exposures, and computed the
median over all objects on a plate, giving results like those on the
left-hand side of Figure~\ref{fig:ratios}. We fit this ratio to the
results expected from a shifting interference pattern (essentially a
derivative of the shifting component in Figure~\ref{fig:superflat}),
with the only free parameter being the amount of shift, and divided
out this remaining component in each spectrum.  The
right-hand panel of Figure~\ref{fig:ratios} shows that this technique
removes the majority of the effects of the shifting interference.  An
example is shown in Figure~\ref{fig:wiggle_spectrum}, the spectrum of
an A star observed on a plate where the interference term was
particularly bad.  The shapes of the absorption
lines, especially H$\epsilon$ at 3970\AA, is much more regular in the
new reductions.

\subsection{Wavelength Calibration}
\label{sec:lambda_cal}

  The spectroscopic wavelength calibration is done quite accurately in
  SDSS, with typical errors of 2 \kms\ or better.  As the DR6 paper
  describes, however, detailed analyses of stellar spectra revealed
  occasional errors that were substantially larger than this,
  especially in the blue end of the spectrum.  The algorithms for
  fitting arc and sky lines were made more robust for DR6, and this
  improved the situation considerably.  We have implemented two  
  further improvements for DR7: 
\begin{itemize} 
  \item Spectroscopy is often done on nights with a moderate amount of
    moon.  The bluest sky line used for wavelength calibration is a Hg
    line at
    4046\AA, which is very close to a strong Fe I absorption
    line in the solar spectrum.  Thus when there is
    substantial moonlight in the sky spectrum, a fit to what is
    assumed to be an isolated emission line can be significantly biased, systematically skewing the
    wavelength solution at the blue end by as much as 20 \kms.  In DR7, we now fit this line
    to a linear combination of a Gaussian plus a stellar template
    including the absorption line, giving an unbiased estimate of the
    wavelength of the line.  	In practice, bright moon affected
    10 plates (listed in Yanny \etal\ 2009) out of a total of 410
    SEGUE plates.  
   \item The sky and arc lines for each fiber are fit to a wavelength
     solution; this is done independently for each fiber.
     This works well for the vast majority of plates.  However, for
     a small fraction of plates, the arcs are weak (perhaps because
     the arc lamps 
     themselves were faulty at that time, or because the petals which
     reflect the arc lamp light
     were not properly deployed), and the wavelength solution
     is poorly constrained.  We therefore required that second- and
     higher-order terms in the wavelength solution be continuous
     functions of fiber number, to constrain the solution.  We found
     that this produces much more robust wavelength solutions  for
     those plates with weak arc observations, and has no substantial
     effect on the remaining plates. 
\end{itemize}

The stellar spectral template library which gives the best 
radial velocity estimates is 
based on the ELODIE library (Prugniel \& Soubiran 2001).  
We have removed one ELODIE template
that gave velocities with a consistent offset from the rest of the
library, as measured using the sample of $\sim 5000$ stars with duplicate
observations on each SEGUE plate pair.  In order to provide more
complete coverage in effective temperature, surface gravity and
metallicity for hot stars, we 
generated a grid of synthetic spectra using the models from Castelli
\& Kurucz (2003) over the same wavelength range
and at the same resolving power as the spectra in the ELODIE library.
This blue grid spans 6000--9500K in 500K increments, $-0.5 > \rm
[Fe/H] > -2.5$
in increments of 0.5 dex, and $\log g$ of 2 and 4.  We also added a grid
of synthetic carbon enhanced spectra (Plez, private communication,
using the stellar atmospheric code described by Gustafsson \etal\ 2008)
at values of [Fe/H] between $-1$
and $-4$, [C/Fe] between 1 and 4, $\log g$ values between 2 and 4, and
$T_{eff}$ in the range 4000K--6000K.  With these improvements, the
radial velocity scatter in repeat observations for 
objects that match the Carbon star templates is now the same as for
the full sample.  

  The DR6 paper describes a 7 \kms\ systematic error in the radial
  velocities of stars (in the sense that the pipeline-reported
  velocities are too small).  This is still with us in DR7; a correction is
  put into the outputs of the SEGUE Stellar Parameter Pipeline (Lee
  \etal\ 2008a) but not elsewhere in the CAS or DAS.  Beyond this
  problem, the plate-to-plate velocities of SEGUE stars have
  systematic errors of about 2 \kms\ in the mean.  The rms velocity error of any
  given SEGUE star observation is about 5.5 \kms\ at $g=18.5$,
  degrading to 12 \kms\ at $g=19.5$.

\subsection{Strong Unresolved Emission Lines}
\begin{figure}\centering\includegraphics[width=12cm]{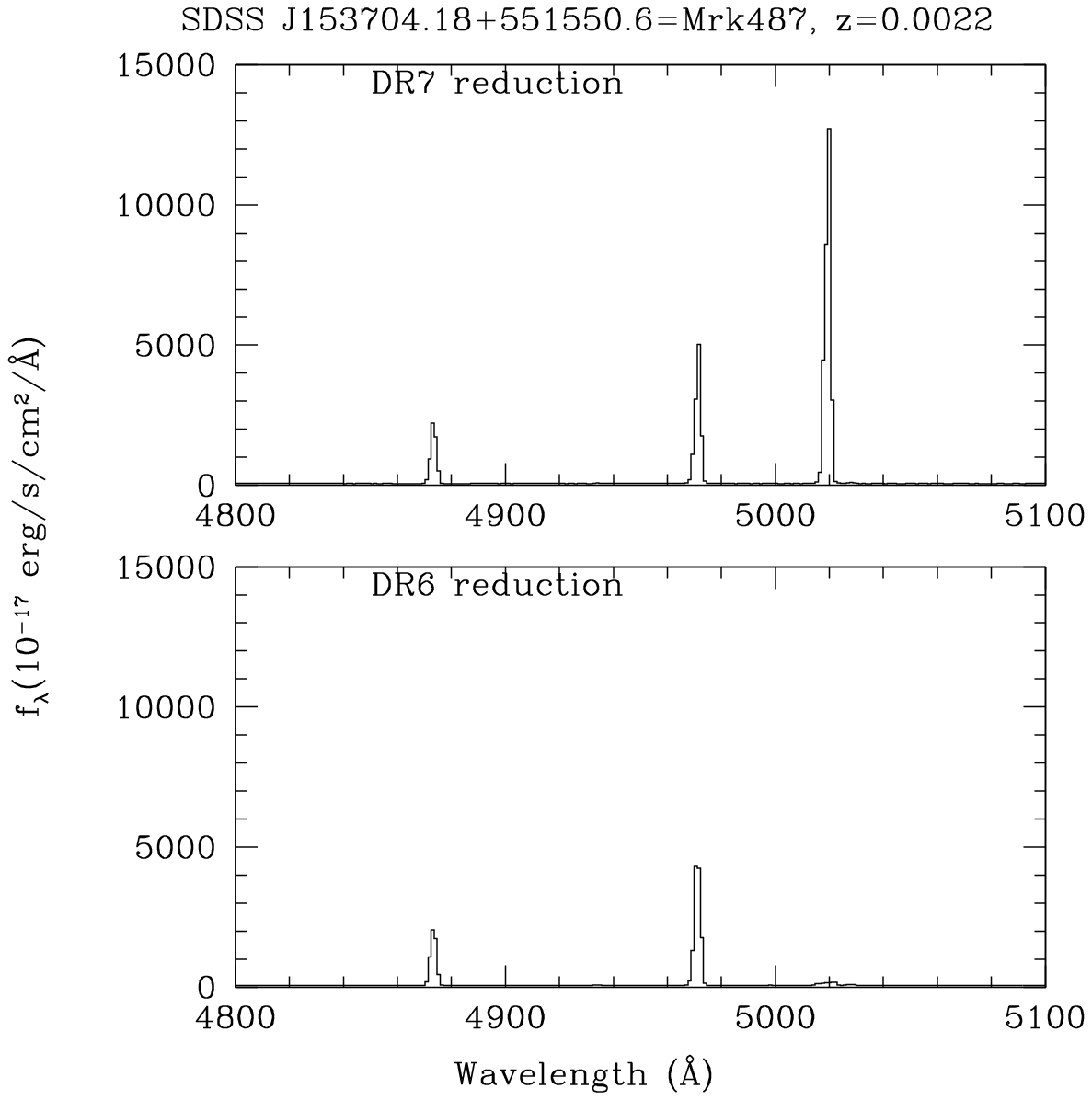}\caption{
Spectra of the object SDSS J153704.18+551550.6=Mrk487, in DR6 and
DR7.  The stronger [OIII] emission line at 5020\AA\ was mistaken for a
cosmic ray and clipped away completely in DR6, while the weaker line
at 4970\AA\ was slightly affected.  With the improved algorithm in
DR7, the lines are not clipped. 
\label{fig:clipped_lines}}\end{figure}

  The spectroscopic pipeline combines observations of a given object
  on the red and blue spectrographs, and between the separate
  15-minute exposures on the sky, by fitting a tightly-constrained
  spline to the data, allowing discrepant points such as cosmic rays
  to be rejected.  This spline requires as input the effective resolution of the spectra.  As
  described in the DR6 paper, it did not do a perfect job;
  occasionally, very strong and sharp emission lines were erroneously
  rejected by this algorithm.  This turned out to be due to the fact
  that the spline code did not adequately track the changing
  resolution of the spectra as a function of wavelength and fiber
  number.  Including this effect significantly improved the behavior of this
  algorithm.  Figure~\ref{fig:clipped_lines} shows an example spectrum of an
  object affected by this problem in DR6, and its improved counterpart
  in DR7, as is apparent by the correct 3:1 ratio of the 5007\AA\ and 4959\AA\
  lines of [OIII].  

  There is another problem, unfortunately not fixed in DR7, which has
  a similar effect.  If the line is so bright that it is saturated in
  the individual 15-minute exposures of the spectrograph, it will also
  appear clipped.  The flux value corresponding to saturation is a
  function of wavelength, but ranges from 2000 to 10,000 times
  $10^{-17}\,\rm erg\,s^{-1}\,cm^{-2}\,\AA^{-1}$ (the units in which
  spectral flux density in reported in the SDSS outputs).
  Unfortunately, such saturated pixels are not 
  flagged as such, although usually they are recognizable as
  having an inverse variance equal to zero.  
  Luckily, objects with such strong emission lines are very rare, but
  the user should be aware of the possibility of objects with
  extremely strong emission lines and unphysical or unusual line
  ratios.  

\subsection{Improvements in the SEGUE Stellar Parameter Pipeline}

There have been several improvements made in
the SEGUE Stellar Parameter Pipeline (SSPP; Lee \etal\ 2008a; 2008b;
Allende Prieto \etal\ 2008a) since the release of DR6. In particular,
in DR6, the SSPP under-estimated metallicities (by about 0.3
dex) for stars approaching solar metallicity.  This was fixed in DR7 by adding synthetic
spectra with super-solar metallicities to two of the synthetic grid matching
techniques ({\tt NGS1} and {\tt NGS2}), and by 
recalibrating the {\tt CaIIK2}, {\tt ACF}, {\tt CaIIT}, and {\tt
  ANNRR} methods. See Table 5 in Lee \etal\ (2008a) for the naming
convention for each technique. Two new techniques ({\tt ANNRR} and
{\tt CaIIK3}) were also added to the SSPP metallicity estimation
schemes, and contributed to the high-metallicity performance
improvement.

Two methods, {\tt ACF} and {\tt CaIIT}, have been recalibrated to the
``native" $g-r$ system, instead of making use of calibration on $B-V$,
which required application of an uncertain transformation in color
space. The {\tt ANNRR} approach, which also tended to under-estimate
metallicity for near-solar metallicity stars, has been re-trained on
the SDSS/SEGUE spectra with improved stellar parameters, resulting in
a better determination of the metallicity for metal-rich
stars. Moreover, a neural network approach, based solely on
noise-added synthetic spectra, has also been introduced. There remains
a tendency for the SSPP to assign slightly higher metallicities for stars
with [Fe/H] $< -2.7$.  This offset is presently being calibrated out, and
will be corrected in SEGUE-2; see below. For more detailed descriptions
of individual methods of the SSPP, we refer the interested reader to
Lee \etal\ (2008a).

Additionally, the pipeline now identifies cool main sequence stars
 of low metallicity (late-K and M subdwarfs). The stars are assigned
 metallicity classes and spectral subtypes following the classification
 system of L\'epine et al. (2007). Cool and ultra-cool subdwarfs are
 classified as subdwarfs (sdK, sdM), extreme subdwarfs (esdK, esdM),
 and ultrasubdwarfs (usdK, usdM) in order of decreasing metal
 content. The classification is based on the absolute and relative
 values of the TiO and CaH molecular bandstrengths, and derived from
 fits to K-M dwarf and K-M subdwarf spectral templates.

A number of open and globular clusters have been observed
photometrically and spectroscopically with the SDSS instruments to
evaluate the performance of the SSPP (Lee \etal\ 2008b).  In addition,
high-resolution spectra have been obtained for about 100 field stars
included in the SDSS, and used to expand the SSPP checks over a wider
parameter space (Allende Prieto \etal\ 2008a).

\section{Looking Ahead to SDSS-III}
\label{sec:conclusions}

This paper marks the release of the final data of SDSS-II.  The
original SDSS science goals (York \etal\ 2000) included five-band
imaging over $10^4$ deg$^2$ with 2\% rms errors or better in
photometric calibration, and spectroscopy of $10^6$ galaxies and
$10^5$ quasars.  We have met these goals, and have in addition carried
out extensive stellar spectroscopy of close to half a million stars,
and repeat imaging over 250 deg$^2$ to search for supernovae.  
Over 2200
refereed papers have been published to date using SDSS data or
results, on subjects 
ranging from the large-scale distribution of galaxies to distant
quasars to substructure in the Galactic halo to surveys of white
dwarfs to the color distribution of main belt asteroids. 

  The SDSS telescope has started a new operational phase, called
  SDSS-III, which will include four surveys with the 2.5m telescope
  through 2014:
\begin{itemize} 
\item SEGUE-2 extends the science goals of SEGUE with the same
instrumentation and data processing pipelines, but targets fainter
stars to study the distant halo.  It will increase the number of
distant halo stars by a factor of 2.5 with respect to the results of
SDSS and SDSS-II.
\item The Baryon Oscillation Spectroscopic Survey (BOSS) will perform 
  spectroscopy of 1.5 million luminous red galaxies to $z \approx 0.7$
  and 160,000 quasars with $2.3 < z < 3$ to measure the scale of the
  baryon oscillation signal in the correlation function as a function
  of redshift (Schlegel \etal\ 2007). 
\item The Multi-object APO Radial Velocity Exoplanet Large-area Survey
  (MARVELS) will monitor the radial velocities of 11,000 bright stars
  to search for the signature of planets with periods ranging from
  several hours to two years (Ge \etal\ 2008).
\item The APO Galactic Evolution Experiment
  (APOGEE) will perform $R \approx 20,000$ $H$-band spectroscopy of
  $10^5$ giant stars to $H=13.5$ for detailed radial velocity and
  chemical studies of the Milky Way (Majewski \etal\ 2008; Allende
  Prieto \etal\ 2008b). 
\end{itemize}

These data will be made public in a series of data releases,
following the pattern established by SDSS 
and SDSS-II.  

\acknowledgements

This paper represents the end of SDSS-II, the culmination of a
project taking two decades and involving an enormous number of
scientists from all over the world.  We would like to dedicate this
paper to colleagues who made essential contributions to the SDSS but
are no longer with us: 
John N. Bahcall, Don Baldwin, Norm Cole, Arthur Davidsen, Jim Gray,
Bohdan Paczy\'nski, and David N. Schramm.   
The successful completion of this project is in large part a
reflection of the hard work and intellectual capital they put into
it.  

Funding for the SDSS and SDSS-II has been provided by the Alfred P. Sloan 
Foundation, the Participating Institutions, the National Science Foundation, 
the U.S. Department of Energy, the National Aeronautics and Space 
Administration, the Japanese Monbukagakusho, the Max Planck Society, 
and the Higher Education Funding Council for England. The SDSS Web Site 
is \url{{\tt http://www.sdss.org/}}.

The SDSS is managed by the Astrophysical Research Consortium for the 
Participating Institutions. The Participating Institutions are the 
American Museum of Natural History, Astrophysical Institute Potsdam, 
University of Basel, University of Cambridge, Case Western Reserve University, 
University of Chicago, Drexel University, Fermilab, the Institute for 
Advanced Study, the Japan Participation Group, Johns Hopkins University, 
the Joint Institute for Nuclear Astrophysics, the Kavli Institute for 
Particle Astrophysics and Cosmology, the Korean Scientist Group, 
the Chinese Academy of Sciences (LAMOST), Los Alamos National Laboratory, 
the Max-Planck-Institute for Astronomy (MPIA), the Max-Planck-Institute 
for Astrophysics (MPA), New Mexico State University, Ohio State University, 
University of Pittsburgh, University of Portsmouth, Princeton University, 
the United States Naval Observatory, and the University of Washington.

\end{document}